\documentclass[authoryear,final,3p]{elsarticle}

\usepackage{amssymb}
\usepackage{amsmath}
\usepackage{lineno}
\usepackage{xcolor}
\usepackage{tabularray}
\usepackage{rotating}
\usepackage{lipsum}
\usepackage{mathtools}
\usepackage{cuted}
\usepackage{physics}
 \usepackage{colortbl}
 \usepackage[normalem]{ulem}
\useunder{\uline}{\ul}{}
\newcolumntype{L}{>{\arraybackslash}m{3in}}

\definecolor{ceruleanblue}{rgb}{0.16, 0.32, 0.75}

\usepackage{cite}
\usepackage[colorlinks=true,linkcolor=blue,allcolors=blue]{hyperref}%
\usepackage[authoryear]{natbib}

\usepackage{epsfig,color,amsmath}
 
\usepackage{wrapfig}
\usepackage{setspace}  
\usepackage{xcolor}
\usepackage{epstopdf}
\usepackage{amssymb}
\usepackage{url}
\usepackage{mathtools}
\usepackage{enumitem}
\usepackage{multirow}
\usepackage{makecell}
\usepackage{amsmath}
\usepackage{stackrel}
\usepackage{booktabs}
\usepackage[flushleft]{threeparttable}
\usepackage{makecell}

\usepackage{subfig}
\usepackage{graphicx}
\usepackage[normalem]{ulem}
\useunder{\uline}{\ul}{}
\usepackage{lscape}

\usepackage[linesnumbered,lined,boxed,commentsnumbered,ruled,longend]{algorithm2e}


\setlist[itemize]{leftmargin=*}

\makeatother

\newcommand{\cs}{{\cal s}}

\newcommand{\m}{\boldsymbol}
\allowdisplaybreaks[4]
\newcommand{\mr}[1]{\mathrm{#1}}

\usepackage{amsthm}

\usepackage[usestackEOL]{stackengine}
\stackMath
\usepackage[framemethod=TikZ]{mdframed}
\usepackage{bm}
\mdfdefinestyle{MyFrame}{%
	linecolor=black,
	outerlinewidth=1.5pt,
	roundcorner=1.5pt,
	innerrightmargin=5pt,
	innerleftmargin=5pt,}

\usepackage{tikz}
\usetikzlibrary{matrix,positioning,decorations.pathreplacing}

\usepackage[export]{adjustbox}
\usepackage{lipsum}
\usepackage{mathtools}
\usepackage{cuted}
\usepackage{physics}
 \usepackage{colortbl}
 \usepackage[normalem]{ulem}
 \usepackage{float}
\useunder{\uline}{\ul}{}

\definecolor{ceruleanblue}{rgb}{0.16, 0.32, 0.75}

\journal{Elsevier}

\begin{document}

\begin{frontmatter}



% \title{Spatio-Temporal Performance of 2D Local Inertial Hydrodynamic Numerical Schemes: Investigating the Effects of Urban Drainage Infrastructure and the Applicability for Dam-Break Scenarios}

% \title{Spatio-Temporal Performance of 2D Local Inertial Hydrodynamic Numerical Schemes: Investigating the Effects of Urban Drainage Infrastructure and the Applicability for Dam-Break Scenarios}

\title{Spatio-Temporal Performance of 2D Local Inertial Hydrodynamic Models for Urban Drainage and Dam-Break Applications}

% Spatio-Temporal Skill Performance Assessment of Local-Inertial 2D Model Numerical Schemes for Scenarios of Dam-Break and Urban Catchments


\affiliation[1]{organization={University of Arizona},
	addressline={Department of Hydrology and Atmospheric Sciences, James E. Rogers Way, 316A}, 
	city={Tucson},
	%          citysep={}, % Uncomment if no comma needed between city and postcode
	postcode={85719}, 
	state={Arizona},
	country={United States of America}}  

\affiliation[2]{organization={University of São Paulo, Department of Hydraulic Engineering and Sanitation, São Carlos School of Engineering},
           addressline={Av. Trab. São Carlense, 400 - Centro}, 
           city={São Carlos},
        citysep={}, % Uncomment if no comma needed between city and postcode
          postcode={13566-590}, 
           state={São Paulo},
           country={Brazil}}
\affiliation[3]{organization={Institute of Hydraulic Research, Federal University of Rio Grande do Sul},
            addressline={Av. Bento Gonçalves, 9500}, 
            city={Porto Alegre},
          citysep={}, % Uncomment if no comma needed between city and postcode
            postcode={91501-970}, 
            state={Rio Grande do Sul},
            country={Brazil}}     
            
\affiliation[4]{organization={The University of Texas at San Antonio, College of Engineering and Integrated Design, School of Civil \& Environmental Engineering and Construction Management},
            addressline={One UTSA Circle, BSE 1.310}, 
            city={San Antonio},
          citysep={}, % Uncomment if no comma needed between city and postcode
            postcode={78249}, 
            state={Texas},
            country={United States of America}}  

\affiliation[5]{organization={University of Bristol, School of Geographical Sciences},           
            city={Bristol},
          citysep={}, % Uncomment if no comma needed between city and postcode
            postcode={BS8 1SS}, 
            country={United Kingdon}}  

\affiliation[6]{organization={Fathom},           
            city={Bristol},
            addressline={7-18 Berkeley},
          citysep={}, % Uncomment if no comma needed between city and postcode
            postcode={BS8 1HB}, 
            country={United Kingdon}}

% Authors
\author[1,4]{Marcus N. {Gomes Jr.}
        }
\author[2]{Maria A. R. A. Castro 
        }
 \author[2]{Luis M. R. Castillo 
        }  
\author[2]{Mateo H. Sánchez
        }
\author[4]{Marcio H. Giacomoni
        }
\author[3]{Rodrigo C. D. de Paiva}   
\author[5,6]{Paul D. Bates
        }

\begin{abstract}
Accurate and rapid flood modeling is key for flood analysis and forecasting. Full momentum hydrodynamic models can present longer computational times, sometimes even larger than the forecast horizon. Low-complexity models, such as the local-inertial approximations, can provide accurate results under subcritical flow regimes with hypothesized limited performance for supercritical flows. This paper explores two main issues: (i) the role of urban infrastructure in 2D hydrodynamic modeling in areas with the absence of detailed sewer and urban drainage data, and (ii) the accuracy of the 2D local-inertial modeling using three different numerical schemes (original formulation, s-centered, and s-upwind) for a dam-break scenario under a complex and relatively flat terrain. The developed model (HydroPol2D) is benchmarked with HEC-RAS 2D full momentum solver for performance comparison. One numerical case study (i.e., the non-breaking wave on a flat surface) and three real-world case studies, including a detention pond receiving a 1 in 100  year inflow hydrograph, a highly urbanized catchment subjected to a 1 in a 50 year hyetograph (both in São Paulo, Brazil), and a dam-break that would impact a coastal city with nearly 200,000 people that could be affected, are tested. Results indicate that the internal boundary conditions implemented in the model can accurately simulate (within less than 5\% peak errors) culverts and spillways compared to HEC-RAS 2D. By not including the urban infrastructure, peak discharges at the outlet can have approximately 17.5\% difference, and flow hydrographs at the boundary conditions are completely mismatched, while computational times nearly double. The results of the dam-break scenario show that the model can achieve good performance in predicting wet areas for maximum flood depths (Critical Success Index (CSI) = 0.95 [original local-inertial model], 0.92 [s-centered], and 0.89[s-upwind]); however, the lack of convective inertia causes a faster flood wave advance than the full momentum solver. The model can be easily adapted for simulating partial or total dam collapses in forecasting systems, especially because of the 23 times faster computational time required for the application in comparison with the HEC-RAS 2D solver used for Benchmark. Furthermore, a spatial simulation of urban drainage infrastructure, such as orifices, weirs, and pumps controlling flows in channels, reservoirs, and tanks, is implemented, allowing a wider application of the model in urban areas.
\end{abstract}


%\item We develop a modeling framework to assess dam-break scenarios
%\item The model was applied to xx dams in Brazil
%\item The potential hazard was classified into low, medium, and high
%\item  
%\item 


\end{frontmatter}



\newpage
\noindent \textbf{Graphical Abstract}

\begin{figure}
\centering
\includegraphics[width=1\linewidth]
{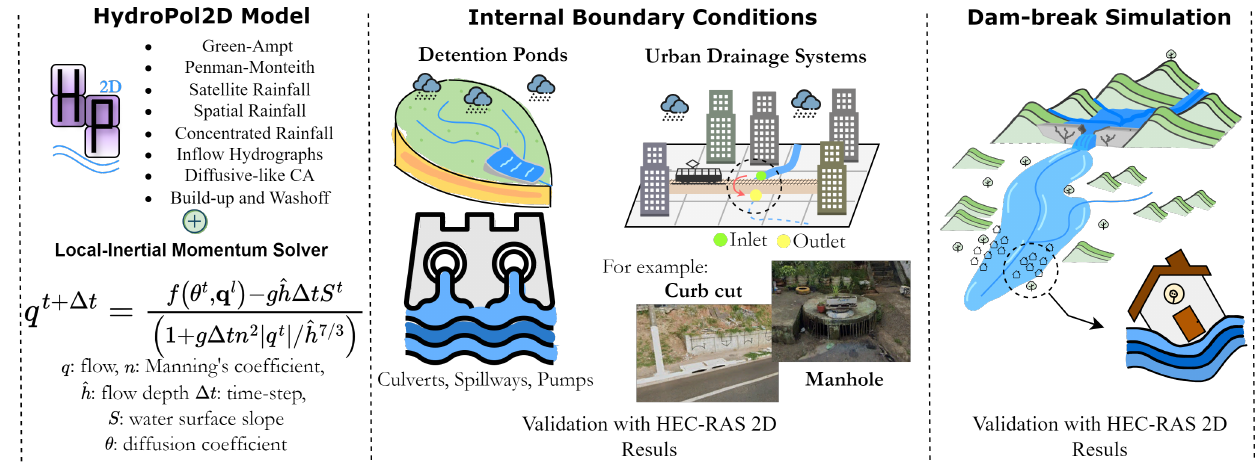}
\end{figure}

\singlespacing     
\section{Introduction}
% Introduction 
Flooding presents the most significant disruption among all natural disasters experienced by human societies \citep{jongman2012global}. Globally, the value of infrastructure exposed to flooding is likely to total tens of thousands of US trillion dollars \citep{jongman2012global}. Flood inundation, hazard estimation, and risk mapping are the primary measures for flood management \citep{mohanty2020flood} requiring skillful models to predict flood behavior and its impacts. Accurate and rapid flood inundation mapping is crucial for flood warning systems \citep{merwade2008uncertainty}. The balance between model complexity reduction and computational and hydrodynamic performance has been extensively researched \citep{afshari2018comparison,bates2010simple,bates2022flood,neal2012much}, with findings converging to simplified, yet accurate, approximations of the complete shallow water equations to some extent, especially because real-world applications in large domains often requires prohibitive computational times for full momentum models \citep{do2023generalizing}.

Full hydrodynamic models account for all terms of the equation for conservation of momentum, including local and convective inertia and gravitational, frictional, and pressure forces \citep{getirana2013mapping}. Neglecting all terms except friction and gravity leads to the kinematic wave model, which is an approximate solution for flows under relatively steep terrain \citep{de2000physically}. Solutions of this approach have been widely applied due to its computational simplicity and reduced simulation times. Several limitations, however, arise from this approach, such as the lack of capability to simulate backwater effects of floodplain resistance connection between inbank and overbank areas \citep{getirana2013mapping}.

The approximation becomes the diffusive or non-inertial wave by including the pressure term in the kinematic-wave approach \citep{hunter2005adaptive}. This simplification is extensively applied in commercial models such as GSSHA \citep{downer2004gssha}, HEC-RAS, HEC-HMS \citep{Brunner}, in early versions of LISFLOOD-FP \citep{hunter2007simple}, and others and typically have good accuracy in coarse grids. Advances from the kinematic wave approach are noted by the possibility of simulating hydraulic transients, backwater effects, and time-varying flow direction that changes according to water surface elevation gradients \citep{gomes2023hydropol2d}, a feature absent in kinematic-wave models \citep{gomes2024real}. Although advances have been made, the main drawback of the diffusive-wave approach occurs for low-relief water surface gradients, frequently causing chequerboard effects (i.e., water moves in cycles around neighbors within consecutive time-steps), decreasing the accuracy of the solution \citep{hunter2005adaptive}. This can happen in reservoirs or coastal areas, for example. To deal with this effect, an adaptive time-stepping scheme can be adopted that resolves the chequerboard issue but requires an impactful reduction in time-stepping, causing the diffusive-wave models (i.e., a theoretically simplified model compared to a full Saint-Venant model) to take longer simulation times than complete full momentum models depending on the catchment physiographic and boundary conditions \citep{bates2010simple}.

The full momentum solver, however, allows unique applications, such as the simulation of hydraulic jumps, changes in flow regime, shocks, and highly variable flows with froude number above the unity. A few examples as HEC-RAS, MIKE3, and LISFLOOD-FP, \citep{george2015dam, phyo2023managing, patel2017assessment,sharifian2022lisflood,tschiedel2018uncertainty} illustrate that these models can provide results with good accuracy in terms of peak flows, peak time, and flood area extent in case of 2D simulations. For 2D simulations, however, the relatively long computation time to run relatively large domains with high-resolution (i.e., $\geq$ 2 million cells for running in personal computers) makes the real-time application of these full dynamic models intractable, even with parallel and graphics processing unit (GPU) computing \citep{bates2023fundamental}. Addressing computational time issues, simplified methods such as the local-inertial approach \citep{bates2010simple, yamazaki2013improving, schumann2013first,siqueira2018toward,pontes2017mgb}, which neglects the term of advective inertia of the momentum conservation equation \citep{bates2010simple}, can provide accurate results for gradually variable flows similarly to those obtained from standard commercial models that solve the complete shallow water equations, but with greater computational efficiency \citep{de2013applicability}. This benefit of faster computational times is especially important in high-resolution terrains that require smaller time-steps for numerical stability \citep{dazzi2018local, hu2019improved, sanders2008integration}. The application of local-inertial hydrodynamic models is an appropriate solution for most of the typical fluvial hydrological processes \citep{pontes2017mgb}, where the maintenance of relatively small Froude numbers is expected \citep{getirana2013mapping}. 

Low-complexity flow models are, hence, an alternative for addressing scenarios that require fast and/or multiple simulations, such as (i) ensemble or probabilistic flood analysis \citep{rapalo4703477developing}, (ii) monte carlo simulations \citep{aronica2012probabilistic}, (iii) local and global sensitivity analysis \citep{gomes2024global,aronica2012probabilistic}, (iv) green-infrastructure placement via optimization \citep{lu2022surrogate}, and (v) automatic parameter estimation \citep{gomes2024global,dung2011multi}. The tradeoffs between accuracy and computational performance come at the cost of inaccurate estimation of flood characteristics at highly variable flows \citep{de2013applicability,sridharan2020explicit,tschiedel2020use}. In addition, the estimated flooded areas are typically overestimated for simulations using coarser DEMs \citep{muthusamy2021understanding,reshma2024real} mainly due to the loss of proper representation of channel conveyance, leading to larger flooded areas in the vicinity of the channel. To address this issue, sub-grid approaches that derive hydraulic properties (e.g., flow area, hydraulic radius) in terms of flood depth can be defined, allowing coarser DEM simulations with similar accuracy than high-resolution simulations \citep{neal2012subgrid,nithila2024novel}.

Urban infrastructure features such as houses and buildings also affect water surface dynamics \citep{chang2015novel}. The literature considers techniques such as elevating building areas, increasing walls at the boundary of urban features and allowing inflow through inlet areas, making building a porous area, or increasing Manning's roughness coefficient of these areas to represent urban areas' characteristics \citep{bellos2015comparing}. Rain-on-the-grid simulations suffer from elevation raising techniques, which highly increase topographic gradients, substantially increasing hydrodynamic model computational times. Other hydraulic constraints rather than the accurate representation of buildings play a role in urban flood dynamics, such as combined sewer systems, culverts, detention ponds, and flood control systems \citep{bates2022flood}. To represent micro drainage, 1D sewer models can be solved by coupling the interaction between the 2D computational domain with the topology of the 1D sewer network \citep{chang2015novel}. The challenge of this approach is the data requirements of the sewer network and the increasing number of parameters of the model \citep{getirana2023urban}. Coupling techniques between the 1D sewer network and 2D overland areas are generally used to simulate urban drainage interactions \citep{wang2021urban}. Besides having a detailed network topology  (e.g., it is hardly available in developing countries), this approach increases the number of parameters in the coupled model and the computational time. 

% However, it is a good alternative when dealing with recurring events (i.e., events where the effects of minor drainage are more representative) \citep{wang2021urban}. 

In the absence of such detailed network data, the effect of urban infrastructure can be simulated with simple hydraulic laws constraining discharges and depths at cells to a certain level (e.g., rating curves, reservoir pumping schedule, channel flow regulation, orifices, and spillway standard hydraulic equations). This approach would only require parameters at the internal boundary condition, considerably reducing the number of parameters and user-defined decisions. In addition, these parameters can be derived by geometric relationships (i.e., culvert area) taken via Google Earth data, for instance. The accurate 2D modeling integration of flow regulation internal boundary conditions is important to allow for a proper representation of flows governed by downstream conditions \citep{fleischmann2019modeling}, allow flow continuity in culverts and bridges \citep{chen2018hydraulic}, or to guarantee that underground reservoirs with flow regulation through pumping are correct estimated in the domain \citep{yazdi2019optimal}. 

% Dam break simulations face various challenges regarding uncertainty assessment, usually because they approach hypothetical scenarios due to the lack of observed data to validate \citep{f__c__b__mascarenhas_2002}. These sources of uncertainty address the accurate prediction of the rupture hydrograph due to factors such as the use of complex equations and limited breach progression data \citep{yang_hao__2023, khilha_lee_2019}, numerical instability in decorrence of the restrictions on the time step, topography data \citep{francesca_aureli__2021} and the most suitable simulation method for the case study.

%Recent advances in dam-break modeling have shown significant progress in terms of numerical efficiency with the use of GPU-based solvers that ensure faster simulations \cite{j__fernndezpato__2023}, ..

Regarding low-complexity 2D hydrodynamic solvers, the recent literature has demonstrated good applicability of local-inertial models to typical hydrological conditions and flooding events, maintaining sub-critical flows. Although tested under synthetic cases, research conducted in \citep{de2013applicability} in a real-world catchment concluded that the local inertial model has better accuracy for Froude numbers below 0.5. Similarly, research conducted in \citep{wang2021urban} indicated that root mean square errors larger than 100\% can occur when dealing with cells with Froude numbers larger than the unity. Depending on the inflow or rainfall boundary conditions, a large Froude number would impact only a few domain cells \citep{fleischmann2020trade}. Although sometimes only a few cells would have supercritical flow conditions, the error is propagated downstream by the interactions from these cells with their neighbors, potentially reducing the model's accuracy. Nonetheless, only a few researches evaluated the local-inertial numerical schemes for highly variable flows (e.g., the case of a real-world dam-break) and in catchments where the role of urban hydraulic infrastructure is not negligible \citep{neal2012much,luke2015hydraulic} or when absence of urban infrastructure modeling can wrongly constraint the flow within the catchment, generating inacurate flood dynamics. There is a computational gain when using such types of low-complexity models compared to full momentum models, and understanding the trade-offs in such cases would encourage the further application of such models for time sensive analysis (e.g., real-time flood forecasting, dam-break probabilistic forecasting, real-time control of stormwater systems).

The objective of this paper is twofold: (i) investigate the spatio-temporal skillfulness of a 2D local-inertial model for different numerical schemes and (ii) investigate the effects of including urban drainage infrastructure in the model. To address these objectives, we implement a new hydrodynamic solver in HydroPol2D \citep{gomes2023hydropol2d} adapted to three different numerical schemes of the local-inertial formulation - the original local-inertial formulation herein denoted as (lim) \citep{bates2010simple}, the s-upwind scheme \citep{sridharan2020explicit}, and the s-centered scheme \citep{sridharan2020explicit,de2013applicability}. We investigate the performance of these three different 2D local inertial solvers to predict flood depths for the maximum flood depth and time-varying flood maps. In addition to the 2D hydrodynamic solver, a spatial simulation of urban drainage infrastructure, such as orifices, weirs, and pumps controlling flows in channels, reservoirs, and tanks, is implemented, allowing a wider application of the model in urban areas. These new features enable the simulation of gradually variable flows and also account for infiltration, rain-on-the-grid boundary conditions of satellite-based rainfall, spatially interpolated from rain-gauges rainfall, concentrated rainfall, and spatial entering of inflow hydrographs in inlet cells \citep{gomes2023hydropol2d,gomes2024global}.

More specifically, we provide an extensive spatio-temporal comparison between the local-inertial version of HydroPol2D with the benchmark case of HEC-RAS 2D full momentum model, comparing the performance not only at the maximum flood depths but also temporally during the passage of the flood wave, in addition to the standard case of assessing the maximum flood depths and extent. We apply the model in one theoretical setup and three real-world case studies. The non-breaking wave propagating in a flat surface (i) is tested to assess the model performance against an analytical solution of the 1D shallow water equations \citep{hunter2005adaptive} under conditions where the original local inertial formulation suffers from instability. Following, (ii) a detention pond in a headwater catchment with flow controlled by a culvert and spillway \citep{gomes2024real} is simulated and compared with HEC-RAS 2D. To assess the effects of the urban infrastructure in an urbanized catchment, we compare HydroPol2D modeling results in a (iii) real-world catchment where LiDAR information is available and water is being incorrectly ponded due to inaccurate representation of urban drainage infrastructure. Finally, (iv) the simulation of a Dam-Break scenario in Pirapama Dam - Brazil that is upstream of Cabo Santo Agostinho city, which has more than 200,000 inhabitants that could be affected, is evaluated, identifying the spatio-temporal performance of the model against the benchmark HEC-RAS 2D model for flood depths, inundation extent, and computational time.

\section{Methods} \label{sec:modelo}

\subsection{Local Intertial 2D Hydrodynamic Model}
\begin{figure}
    \centering
    \includegraphics[width=1\linewidth]{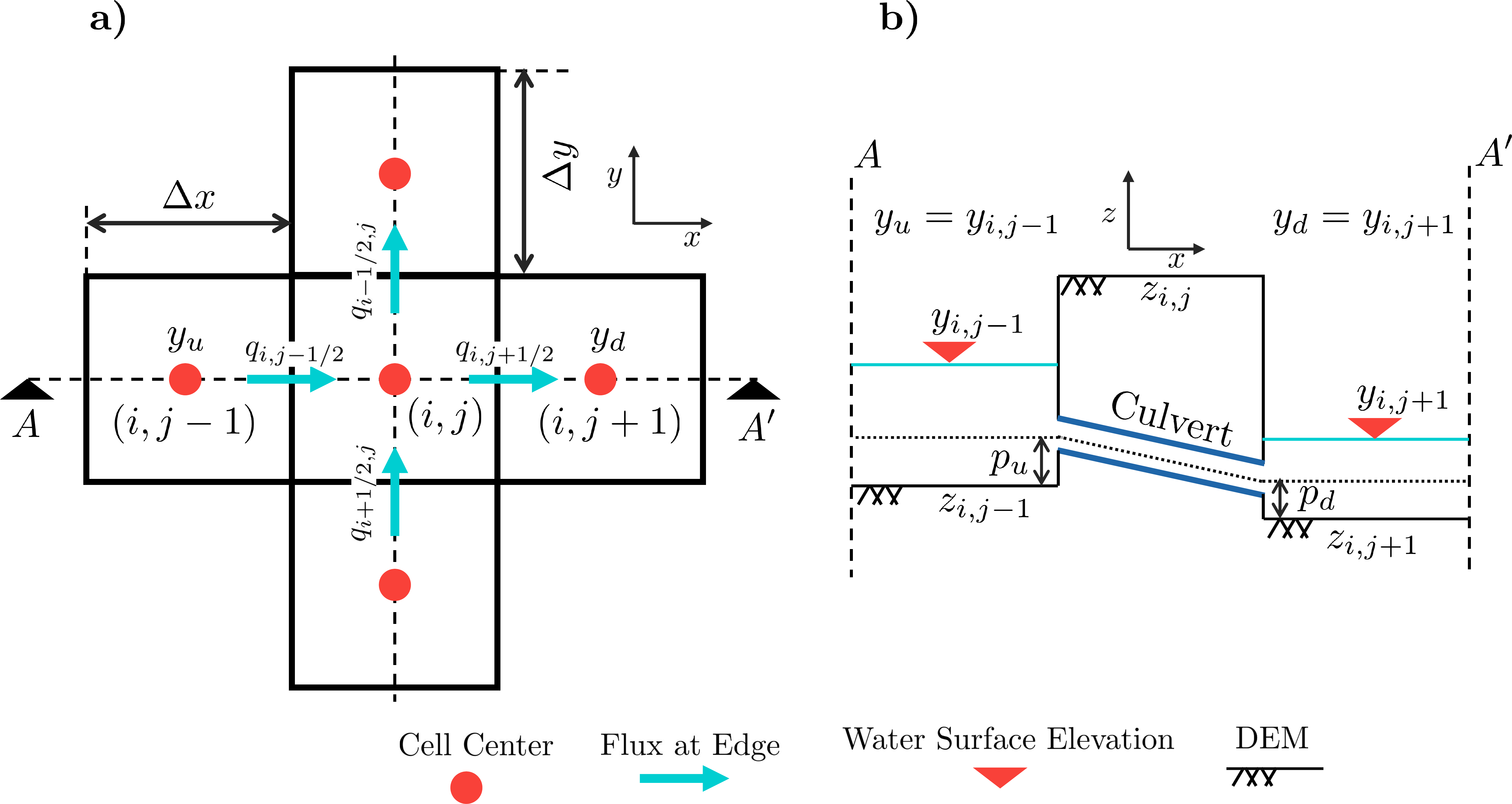}
    \caption{Staggered grid discretization (a) and internal boundary condition treatment (b). The fluxes at the cell edges are given by $q$ [$\mr{L^2 \cdot T^{-1}}$], $\Delta x$ and $\Delta y$ are the grid spatial discretization in $x$ and $y$, respectively [$\mr{L}$], $y$ is the water surface elevation [$\mr{L}$], $z$ is the DEM elevation [$\mr{L}$]. In the example provided in (b), a culvert with depths from the bottom of $p_u$ [$\mr{L}$] and $p_d$ [$\mr{L}$] from upstream (u) and downstream (d) of the boundary condition are represented.}
    \label{fig:grid}
\end{figure}

HydroPol2D solves the momentum equation for cartesian 4D (Von-Neuman) structured grid via a Godunov‐like approach. Matrixwise expressions are solved in the GPU. The model is implemented in MATLAB. The momentum equation from the 1D Saint-Venant Equations accounts for the flow local and advective acceleration and considers the effects of hydrostatic pressure, potential energy, and losses through friction and can be written as:

\begin{equation}
\underbrace{\frac{\partial Q}{\partial t}}_{\text {local inertia }}+\underbrace{\frac{\partial}{\partial y}\left[\frac{Q^2}{A}\right]}_{\text {advective inertia }}+\underbrace{\frac{g A \partial(h+z)}{\partial y}}_{\text {water slope }}+\underbrace{\frac{g n^2 Q^2}{R^{4 / 3} A}}_{\text {friction slope }}= 0
\end{equation}
where $Q$ [$\mr{L \cdot T^{-3}}$] is the flow discharge, $A$ [$\mr{L^2}$] is the cross-sectional area, $z$ [$\mr{L}$] is the topographic elevation, $n$ [$\mr{T \cdot L^{-1/3}}$] is the manning rfoughness, $g$ [$\mr{L \cdot T^{-2}}$] is the gravity acceleration, $R$ [$\mr{L}$] is the hydraulic radius, $t$ [T] is the time, $y$ [L] is longitudinal coordinate of the channel, and $h$ [$\mr{L}$] is the water depth. Fig.~\ref{fig:grid} shows a schematic of the fluxes acting at cell edges in a 2D staggered grid.

For sake of simplicity, only the discretization for $y$ direction is considered. Neglecting the advective terms \citep{hunter2007simple}, using an explicit forward-in-time scheme for the local acceleration, and dividing the two sides of the equation by the width of the grid $\Delta y$, the momentum equation becomes:

\begin{equation} \label{equ:bates}
q^{t+\Delta t}_{i+ 1/2}=q^t_{i + 1/2}-g \hat{h}_{i+ 1/2} \Delta t\left[\frac{\partial\left(h +z\right)}{\partial y}+\frac{n^2 \abs{q^t_{i+ 1/2}}q^t_{i+ 1/2}}{\hat{h}^{10 / 3}_{i  + 1/2}}\right],
\end{equation}
where $q$ [$\mr{L^2 \cdot T^{-1}}$] is the flow discharge per unit width ($Q/\Delta y$), $\Delta y  = \Delta x$ is the grid resolution [L]. The hydraulic radius $R$ can be approximated by the effective flow water depth $\hat{h}$, which can be calculated as \citep{de2013applicability}:

\begin{equation}
\hat{h}_{i + 1/2}=\max \left(h_i+z_i, h_{i+1}+z_{i+1}\right)-\max \left(z^i, z^{i+1}\right)
\end{equation}

From now on, we will drop the notation to indicate the cell interface ($i + 1/2$) in the following equations for notation simplicity. All equations are valid for horizontal and vertical interfaces. The water slope gradient term is discretized at the edges of the cell, such that:

\begin{equation}
    \frac{\partial\left(h+z\right)}{\partial y} =
S^t=\frac{\left(h_{i+1}^t+z_{i+1} - h_i^t - z_i\right)}{\Delta y}
\end{equation}

Using a semi-implicit scheme in the friction term of Eq.~\eqref{equ:bates} by considering $\abs{q^t}q^t \approx \abs{q^t} q^{t + \Delta t}$, the 
momentum equation becomes \citep{bates2010simple}:

\begin{equation} \label{equ:bates_final}
q^{t+\Delta t}=\frac{q^t-g \hat{h} \Delta t S^t }{\left(1+g \Delta t n^2 \abs{q^t} / \hat{h}^{7 / 3}\right)}
\end{equation}

The previous equation has interesting properties as it ensures that small flow depths have limited flow discharges by the effect of $\hat{h}$ in the denominator. The semi-implicit introduction in the friction term makes the equation more stable and still allows for an explicit discretization in the time domain. Eq.~\eqref{equ:bates_final} is solved for each flow direction per cell, in matrixwise operations. 

Wetting and drying interfaces are treated by only allowing fluxes if $\bar{h}_f$ is greater than a threshold $\delta$ and if the available stored depth in the time-step is also greater than $\delta$. Example of this arbitrary value are 0.0001 mm \citep{kim2012coupled,camporese2010surface}, and 1 mm \citep{jones2008application}. Herein, the value of 1 mm is assumed in all simulations. 

Although being a stable equation for subcritical regimes, especially with Froude numbers below 0.5 \citep{de2013applicability}, Eq.~\eqref{equ:bates_final} has reduced accuracy for highly-variable flows, low values of surface roughness (i.e., $n \leq 0.03~\mr{s \cdot m^{-1/3}}$), and flat areas \citep{de2013applicability}. A few improvements can be made in Eq.~\eqref{equ:bates_final} by introducing an artificial numerical diffusion in the flux term of the numerator of Eq.~\eqref{equ:bates_final}, such that:

\begin{equation} \label{equ:bates_diffusive}
q^{t+\Delta t}=\frac{f(\theta^t,\m{q}^t) - g\hat{h} \Delta t S^t }{\left(1+g \Delta t n^2 \abs{q^t} / \hat{h}^{7 / 3}\right)}
\end{equation}
where $f(\theta,\m q)$ controls the numerical diffusion added in the computation of fluxes at  cell boundaries, $\m{q}$ is a vector containing the fluxes in all face neighbors of $i+1/2$, and $\theta$ can be estimated by:

\begin{equation}
    \theta^t=1-\frac{\Delta t}{\Delta x} \min \left(\frac{\left|q^t\right|}{\bar{h}^t}, \sqrt{g \hat{h}^t}\right)
\end{equation}

The term $f(\theta^t,~\m{q}^t)$ accounting for the artificial diffuson varies according to the numerical scheme. In HydroPol2D, besides the original formulation proposed by \citep{bates2010simple}, two numerical schemes are allowed by the manipulation of $f(\theta^t,~\m{q}^t)$ in Eq.~\eqref{equ:bates_diffusive}, the s-upwind scheme and the s-centered scheme \citep{sridharan2020explicit}.

\begin{equation}
f(\theta^t,\m q^t) = 
\begin{cases}
    q^t, ~~~~~~~~~~~~~~~~~~~~~~~~~~~~~~~~~~~~~~~~~~~~~~~~~~~~~~~~~~\textit{local inertial model (lim)} \\ 
     \theta^t q^t + (1-\theta^t) \Bigl(\frac{q_{i+3/2,j}^t + q^t_{i - 1/2,j}}{2}\Bigr),~~~~~~~~~~~~~~~~~~~\textit{s-centered scheme}\\
     \begin{cases}
              \theta^t q^t+(1-\theta^t) q_{i-1/2,j}^t,~\text{If } q^t < 0, \\
              \theta^t q^t+(1-\theta^t) q_{i + 3/2,j}^t,~\text{If } q^t \geq 0,~~~~~~~~~~~~~~~~~\textit{s-upwind scheme}
     \end{cases}

\end{cases}
\end{equation}

Both schemes are allowed in the code \citep{de2013applicability,sridharan2020explicit} besides the original local-inertial formulation \citep{bates2010simple}. The adaptive diffusive coefficient $\theta^t$ includes extra computations compared to the original local-inertial formulation; however, increasing courant numbers can be used to counterbalancing the required computational time while increasing model stability \citep{sridharan2020explicit}.

Coupled with the momentum equations, the continuity equation is also solved. The shallow water equations are hence solved in non-conservative form by the inclusion of sink/source terms such as inflow / outflow discharges ($s$) [$\mr{L \cdot T^{-1}}$], rainfall ($r$) [$\mr{L \cdot T^{-1}}$], and soil infiltration ($f$) [$\mr{L \cdot T^{-1}}$]. For a cell $(i,j)$ in the domain, the continuity equation is written as:

\begin{equation} \label{equ:continuity_equation}
h^{t+\Delta t}_{i,j}=h^t_{i,j}+\Delta t \Bigl( \frac{Q_{i-1 / 2, j}^{t+\Delta t}-Q_{i+1 / 2, j}^{t+\Delta t}+Q_{i, j-1 / 2}^{t+\Delta t}-Q_{ i, j+1 / 2}^{t+\Delta t}}{\Delta x^2} \Bigr) + \Delta t \Bigl(r_{i,j}^{t} - f_{i,j}^{t} + s_{i,j}^{t} \Bigr),
\end{equation}
where $Q = q \Delta x$, the notation $x$ and $y$ indicate the cartesian directions, the subindex $i-1/2$, $i+1/2$ denote the $y$ interfaces of cell $i,j$; similarly, $j-1/2$ and $j+1/2$ denote $x$ interfaces of $i,j$ and $\Delta x = \Delta y$ is the raster resolution $[\mr{L}]$. For detailed reference on the other components of HydroPol2D, refer to \citep{gomes2023hydropol2d}.

The previous equations are only valid for non-negative water depths. In the non-conservative shallow water equations formulation, the Courant-Friedrichs-Lewy (CFL) only considering the wave celerity ($c = \sqrt{g \max(h)}$) is a necessary but not sufficient condition to ensure model stability \citep{bates2022flood}. However, by including the sink/source terms, the water depth can become negative if the integration of the boundary condition fluxes and infiltration rates adopts a sufficiently large time-step. To reduce this effect, a try/catch approach in the code is implemented. If the volume difference due to negative water depths is larger than a threshold $\gamma$, the model returns the time step and reduces it by a user-defined factor of reduction. The computational time-step is updated following:

\begin{equation}
\Delta t =
\begin{cases}
    \alpha \Bigl (\frac{\Delta x}{\sqrt{g h}} \Bigr ),~\text{If } V_e \leq \gamma \\
    \\
      \Bigl [\frac{1}{f_r k}\alpha \Bigl (\frac{\Delta x}{\sqrt{g h}} \Bigr )\Bigr],~\text{Elsewhere}
\end{cases}
\end{equation}
where $\alpha$ varies according to floodplain resistence and is suggested by \citep{bates2010simple} values around 0.2 to 0.7. Herein and in all simulations, $\alpha = 0.4$. $V_e$ is the lost volume due to negative depths [$\mathrm{L^3}$] calculated by the summation of negative depths in the domain, $\gamma$ is the volume error tolerance, adopted as 1\% of the inflow discharge. The index $k$ represents the counter of successive time-steps that did not reach the volume tolerance and $f_r$ is a reduction factor, herein adoped as 2.

\subsection{Internal Boundary Conditions}
Several flow controlling structures such as culverts, spillways, and pumps can alter the regular flow regime in computational cells and divert flow to downstream following hydraulic laws. An schematic of the topologic scheme to simulate internal boundary conditions in HydroPol2D is shown in Fig.~\ref{fig:grid}(b). In general, these types of hydraulic devices can be modeled as a state function of the head following a rating curve type as:

\begin{equation} \label{equ:boundary}
\begin{gathered}
    \phi_t = f_t \Bigl[ k_1 \ \Bigl( \abs{(y_u - p_u) - (y_d - p_d)}\Bigr)^ {k_2} \Bigr ]
\end{gathered}
\end{equation}
where $\phi_t$ is the flow discharge [$\mr{L^3 \cdot T^{-1}}$], $f_t$ models the flow direction derived from the water surface elevation gradient [-], $k_1$ is the rating curve coefficient [$\mr{m^{(3 - k_2)} \cdot T^{-1}}$], $y$ is the water surface elevation given by ($z + h$) applied for upstream $u$ and downstream $d$ cells from the boundary condition, and $k_2$ is the rating curve exponent [-].

In the absence of available GIS data for characterizing urban drainage, a HydroPol2D model can be run for a given design storm, and one can identify the points that likely have urban drainage infrastructure by assessing the maps and videos of flood inundation depths from modeling results. Afterward, using Google Earth or locally available data, one can determine the parameters of Eq.~\eqref{equ:boundary} to represent urban drainage infrastructure.

The previous equation allows for simulation inflows and outflows from a particular cell constrained by the boundary condition. Flows can be positive (i.e., leaving the cell) or negative (i.e., entering the cell) according to water surface elevation gradients, according to $f_t$. The previous general equation allows one to simulate spillways ($k_2 = 1.5$), orifices ($k_2 = 0.5$) and pumps with fixed flow ($k_2 = 0$). The boundary outflow is hence considered in the continuity equation Eq.~\eqref{equ:continuity_equation} as a sink/source term for either cells that the flow is diverted to the cells that receive outflow from.

% \subsection{Negative Water Depths}
% Depending on the time step, the solution of Eq.~\eqref{equ:bates_final} can become negative due to the source terms, although the CFL conditions are satisfied. In addition, surplus velocities can also arise in the wet/dry interfaces. To this end, a try/catch approach that attempts to solve the model and checks if a negative depth occurs in the domain is implemented, and in case that occurs, a gradual reduction in the time step is performed. In addition, flow is limited to the critical velocity at the wet/dry interfaces \citep{nithila2024novel}.

\section{Model Applications}

\subsection{Numerical Validation - Non-Breaking Wave Propagation on a Horizontal Plane}
The diffusive analytical solution of the 1D Saint-Venant Equation in a flat surface, neglecting local and convective acceleration, is given by \citep{hunter2005adaptive,de2013applicability}:

\begin{equation}
h(x, t)=\left\{-\frac{7}{3}\left[n^2 u^2(x-u t)\right]\right\}^{3 / 7}
\end{equation}
where $n$ is the Manning's coefficient [$\mr{T \cdot L^{-1/3}}$], $u$ is the wave velocity [$\mr{L \cdot T^{-1}}$], and $t$ is the time [$\mr{T}$]

By assuming $x = 0$, one can define the stage-hydrograph boundary condition, which is considered in the first node of the model, such that:

\begin{equation}
    h(0,t) = \Bigl(\frac{7}{3} n^2 u^3 t \Bigr)^{3/7},
\end{equation}

The previous boundary condition equation is then propagated downstream according to the local inertial model. The original local inertial formulation can present numerical instabilities for $n \leq 0.03~\mr{s\cdot m^{-1/3}}$ \citep{de2013applicability,sridharan2020explicit}. In this analysis, we test the different local inertial numerical schemes allowed in HydroPol2D in comparison to the analytical solution. For this test, the domain is discretized into 32 x 240 cells of 25 x 25 m with $n = 0.005~\mr{s \cdot m^{-1/3}}$ and $u = 0.635~\mr{m \cdot s^{-1}}$, following \citep{sridharan2020explicit}, and the stage-hydrograph boundary condition is fixed in the first column of the domain.

\subsection{Case Study No. 1}
\begin{figure}
    \centering
    \includegraphics[width=1\linewidth]{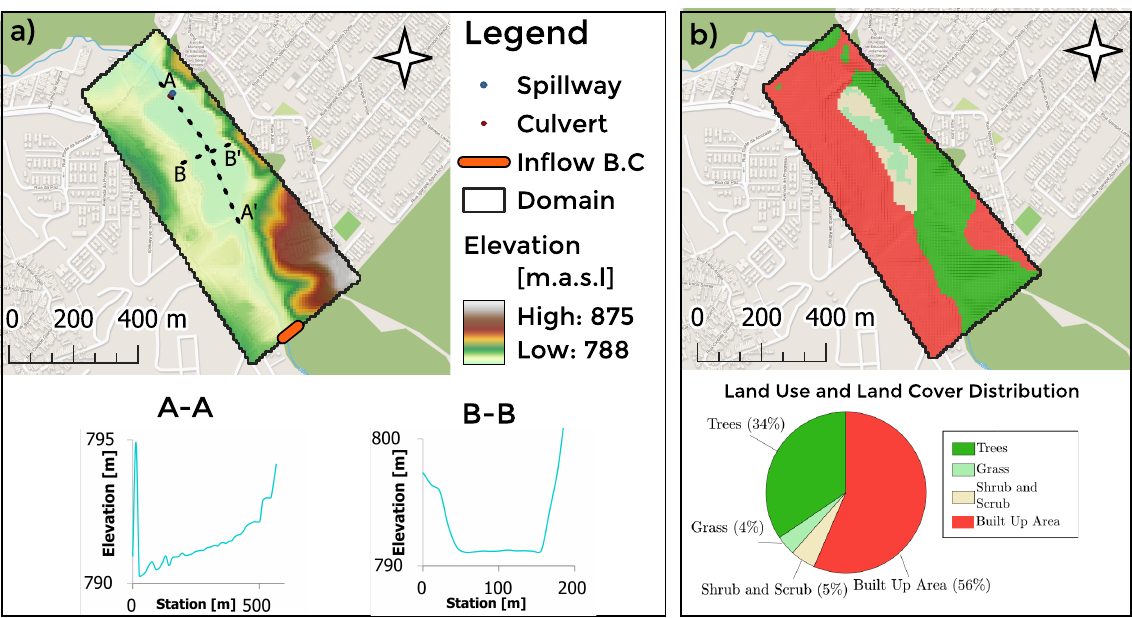}
    \caption{Study area of case study 1, where (a) shows the digital elevation model and (b) the land use and land cover distribution.}
    \label{fig:case_study_ari}
\end{figure}

This case study is centered on the simulation of flow control structures. An online detention pond in the upper part of the Aricanduva  Watershed (AW) \citep{gomes2024real}, situated in the eastern part of São Paulo city, named Aricanduva 1 is considered. In this testing case, only hydrodynamic processes are evaluated; therefore, infiltration is neglected, and only an inflow hydrograph boundary condition and an outlet normal slope boundary condition are considered. A figure of the domain area, DEM, and LULC is shown in \ref{fig:case_study_ari}. Within the domain, internal boundary conditions of the culvert and spillway are imposed, requiring rating-curve ($k_1$) coefficient and exponent ($k_2$) for the box culvert and spillway. These values are 1.82, 1.10 for the culvert and 17.64, 1.5, for the spillway, following Eq.~\eqref{equ:boundary}. These parameters were derived from available plans of the reservoir design, which indicated the culvert dimensions, elevation from the bottom, spillway dimensions, and crest height.

The DEM available at the GeoSampa portal \citep{GeoSampa} was derived from airborne LiDAR DEM survey (1 m resolution) and later resampled to 10-meter resolution. A design inflow hydrograph derived from SCS-CN unit hydrograph using the alternated blocks as the rainfall temporal resolution for a return period of 100 years is tested. In this case, the rainfall duration is assumed to be 80 min, which equals the time of concentration of the upstream catchment. 

The box culvert has a 1.2 x 1.8 meters cross-section and a longitudinal length of 32.4 meters. The Manning's roughness is spatially varied according to the LULC classification (see Supplemental Information for detailed reference). HEC-RAS culvert parametrization requires entrance and exit loss coefficients, and these were assumed to be 0.5 and 1, respectively. A least-square fit of the culvert stage-discharge curve is performed to obtain the representative $k_1$ and $k_2$ values for the univocal rating curve of Eq.~\eqref{equ:boundary}. The spillway has the shape of a broad crest of 4 meters width, a length of 10 meters, a height from the bottom of 2 meters, and a discharge coefficient of 1.66. We compare the reservoir's simulation with a HEC-RAS 2D full momentum solver.

\subsection{Case Study No. 2}
This case study is focused within the upper part of the Franquinho Watershed (FW), located in the eastern part of the São Paulo City, Brazil. The HydroPol2D calibrated parameters for the FW are presented in the Supplementary Information (SI). DEM was derived with airborne LiDAR with 5 m spatial resolution (resampled from 1 m), available at the GeoSampa portal \citep{GeoSampa}. For the Land Use and Land Cover (LULC) with four classes, the "Husqvarna Urban Green Space Index" (HUGSI) data was used \citep{HUGSI}. The FW is mainly located on red-yellow dystrophic clay soil \citep{nachtergaele2023harmonized} and Green-Ampt \citep{green1911studies} soil parameters were calibrated (see Supplemental Material for reference). This duration was choosen due to that most of the design of urban drainage infrastructure in the metropolitan region of São Paulo considered 2-hours as the critical rainfall duration \citep{canholi2015drenagem}.

We compare the base simulation of the urban area without including urban drainage systems devices such as culverts and manholes in the watershed with a simulation where these structures are treated as internal boundary conditions in the model. A Huff hyetograph \citep{huff1967time} with a 2-hour duration and a return period of 50 years is tested, according to IDF curves for the region \citep{Atlas2020}. The computational domain has 94,440 cells, resulting in a drainage area of 2.36 $\mathrm{km^2}$.

\subsection{Case Study No. 3}
In this case, we test a dam break scenario of the Pirapama dam.  The dam is located on the south coast of Pernambuco, Brazil, in the city of Cabo de Santo Agostinho. The Pirapama System is used for human supply and constitutes the main water supply system in the region as it has a greater production capacity, being responsible for 35\% of services in the metropolitan region of Recife, which is the capital of the state. The domain boundaries were defined to minimize the number of cells while allowing a suitable flow area. The domain perimeter has gradient outlet boundary conditions with friction slope assumed as 0.01.

\begin{figure}
    \centering
    \includegraphics[width=1\linewidth]{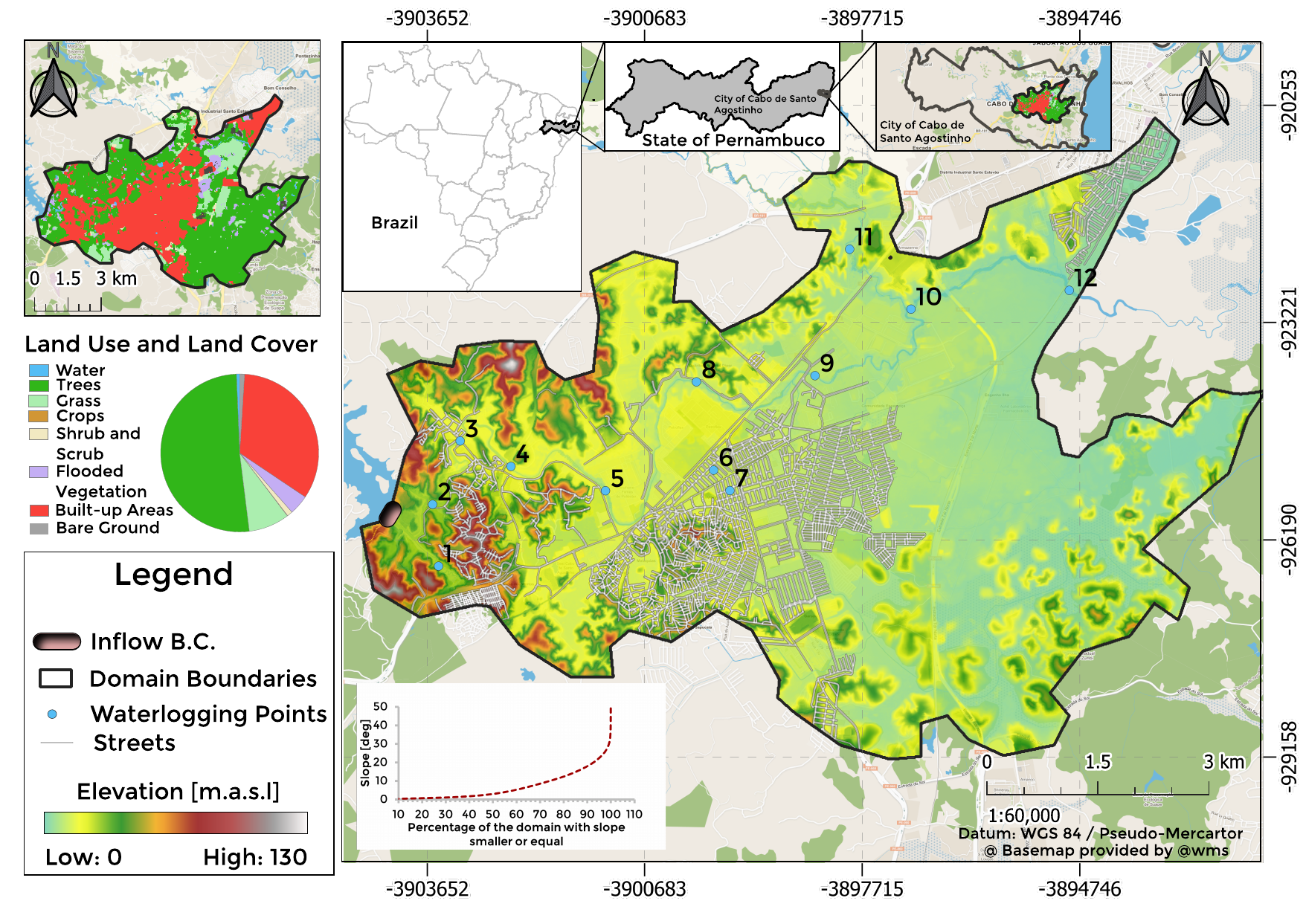}
    \caption{Study area of the Dam 7866 in the City of Cabo de Santo Agostinho showing the land use and land cover maps (LULC), digital elevation map (DEM), and hypsometric slope curve indicating the percentage of areas with slope smaller or equal. Water depths are saved in every waterlogging point.  }
    \label{fig:case_study_dam}
\end{figure}

The heavy rains recorded in 2022 in Pernambuco, which caused the structure to reach its maximum capacity, the proximity of the dam to the city, and the fact that the city is among the 10 most populous in the state, reinforce the need to assess the hazard impacts of a dam failure event. The population of the city of Cabo de Santo Agostinho reached 203,216 people and has a demographic density of 456.27 inhabitants per km² \citep{marengo2023flash}.

The dam is classified as high risk and potential damage by the National Dam Safety Information System (SNISB), which constitutes a consolidated register of information on dams at a national level and takes into account the damage that the dam can potentially cause to elements located downstream, but also the technical characteristics, the state of conservation, and the provision of a safety plan. 

The dam's height is approximately 42 meters, and it can store a volume of 58,5 hm³. As a way of capturing most of the characteristics of the problem with the greatest simplification possible, the dam dimensions will be simplified into a prism of width (B), length (W), and height (h). An instantaneous dam-break hydrograph is used in function of these dimensions \citep{gomes2024increasing}. 

The input data used as digital elevation data was a LiDAR DEM with a 1-meter spatial resolution (resampled for 10 m), available at the PE3D website \citep{PE3D}. For the Land Use and Land Cover (LULC) with eight classes, the Dynamic World data was used \citep{brown2022dynamic}. It was not considered soil classes for this study case due to the limited impact of infiltration in the domain for such inflow hydrograph. The mesh grid accounts for 569,254 cells with an average cell size of 10 m.

% Inflow Hydrograph Figure
The inflow hydrograph is simulated considering an instantaneous dam collapse that can be simulated assuming a rectangular breach as:

\begin{equation}
Q^t =
\begin{cases}
     Q_p, ~~~~~~~~~~~~~~~~~~~~~~~~~~~~~~~~~~~~~\quad 0 \leq t<t_p, \\
    \left[\frac{-1 / 2 h\left(t_p-t\right)(\lambda B \sqrt{8})^{2 / 3}}{S^{t_p}}+Q_p^{-1 / 3}\right]^{-3}, t_p \leq t \leq t_e,
\end{cases}
\end{equation}
where $Q_p$ is the peak discharge calculated by an open crest spillway equation [$\mr{L^3 \cdot T^{-1}}$], $\lambda$ is a discharge coefficient assumed as 8/15 for rectangular breaches, $B$ is the breach width [$\mr{L}$], $t_p$ is the peak time or stable duration where the inflow is maintained constant calculated in terms of the wave celerity and reservoir length [$\mr{T}$], $S^{t_p}$ is the reservoir storage at $t_p$ [$\mr{L^3}$], and $t_e$ is the empty time [$\mr{T}$] \citep{gomes2024increasing}. The resulting hydrograph for this dam is presented in the supplemental material and has a peak discharge of approximately 49.000 $\mr{m^3 \cdot s^{-1}}$ and $t_p = 11~\mr{min}$.

\section{Results and Discussion} \label{sec:discussoes}

\subsection{Non-breaking wave}
\begin{figure}
    \centering
    \includegraphics[width=1\linewidth]{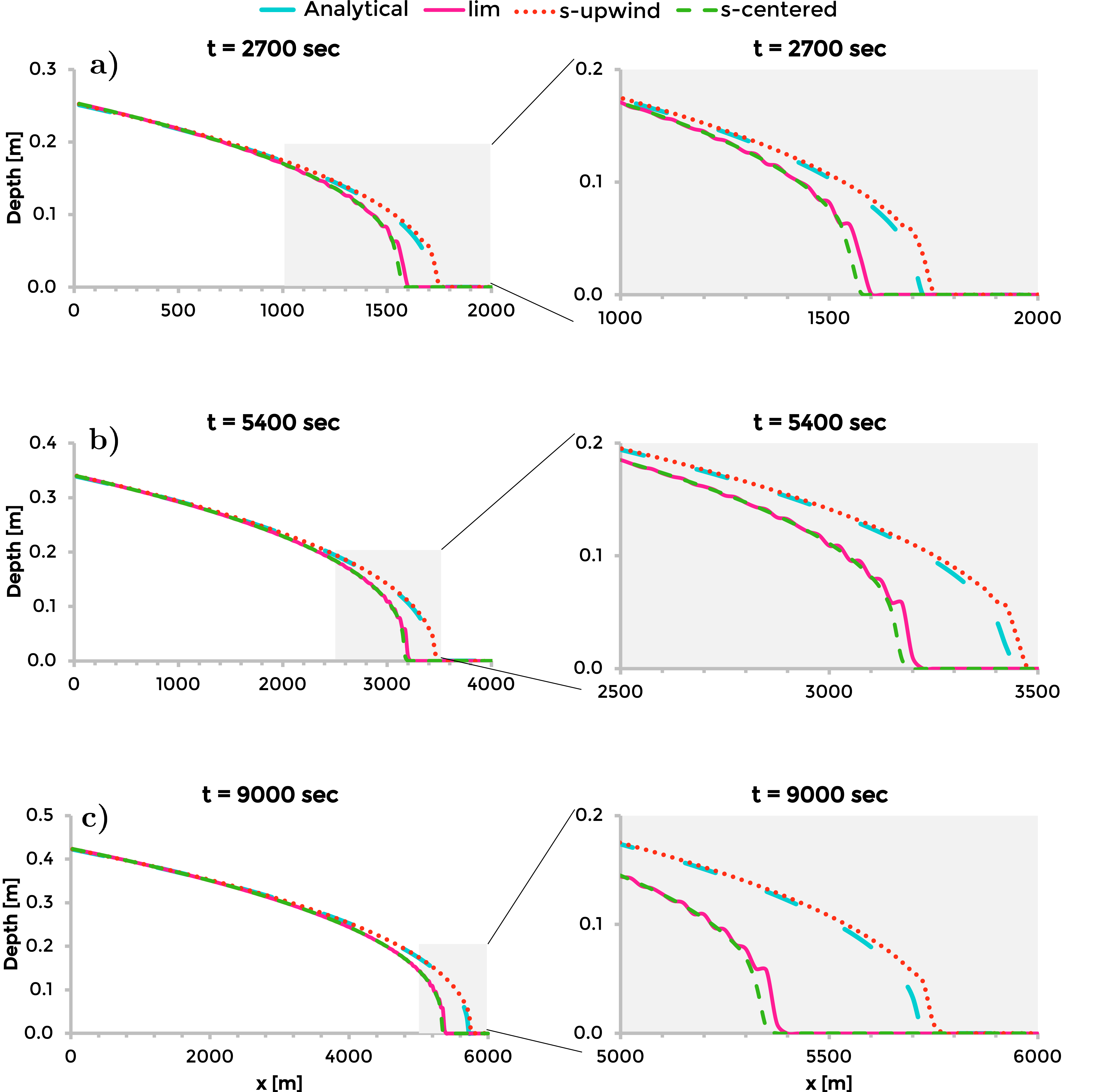}
    \caption{Results of the propagation of a non-breaking wave with $n = 0.005~\mr{s \cdot m^{-1/3}}$, $u = 0.635~\mr{m \cdot s^{-1}}$ for the Local-Inertial-Model (lim) original formulation \citep{bates2010simple}, s-upwind, s-centered, and analytical solution. Part (a) - (c) show results for $t = 2700,~5400,$ and $9000$ sec, respectively.}
    \label{fig:non_breaking_wave}
\end{figure}

The propagation of a non-breaking wave in a flat surface with a relatively low roughness (i.e., $n = 0.005~\mr{s \cdot m^{-1/3}}$ is presented in Fig.~\ref{fig:non_breaking_wave}. The HydroPol2D results are exactly the ones predicted by \citep{de2013applicability} for the s-centered scheme and by \citep{sridharan2020explicit} for the s-upwind scheme (RMSE = 0). The original local inertial formulation of \citep{bates2010simple} presents some instability near the wetting/drying front as shown in the zoomed charts of Fig.~\ref{fig:non_breaking_wave}. The introduction of numerical diffusion by the s-centered scheme, although it produces stable solutions, has larger deviations from the analytical solution. The s-upwind scheme had a closer agreement with the analytical solution for a case of flat and smooth surface \citep{de2013applicability}. Although good agreements between the analytical solutions and the s-upwind scheme were found, the shallow water equation analytical solution derived in \citep{hunter2005adaptive} only considers diffusive effects in the SWE.

\subsection{Case Study No. 1}
\begin{figure}
    \centering
    \includegraphics[width=1\linewidth]{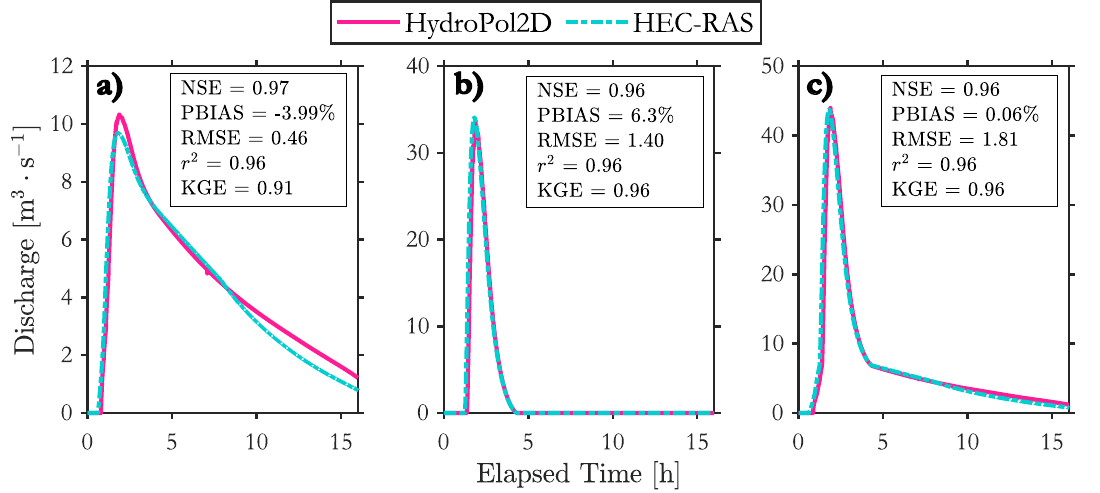}
    \caption{Comparison between HydroPol2D and HEC-RAS discharge simulation under an unsteady-state inflow hydrograph representing the 1 in a 100 years return period. Part (a) is the discharge at the culvert cell, (b) is at the spillway cell, and (c) at the domain outlet.}
    \label{fig:boundary}
\end{figure}

% Discharges comparing Inertial, CA, and HEC-RAS 
The hydrographs comparing HydroPol2D with HEC-RAS 2D for the culvert and spillway are shown in Fig.~\ref{fig:boundary}. The representation of the flow control structures as univocal rating-curves, described as the Eq.~\eqref{equ:boundary}, shows a good performance compared to HEC-RAS 2D solver. HEC-RAS 2D solves the culvert equations by considering downstream and upstream flow conditions and water surface elevations and changes the governing equations according to hydraulic regimes \citep{Brunner}. Therefore, an expected difference in the prediction of hydrographs and, hence, altering the reservoir stage is expected. Fig.~\ref{fig:boundary}a), however, shows a deviation in peak discharges of approximately 5\%, which, in this case, indicates that a simple rating curve relationship can be sufficiently accurate to simulate flow discharge under a relatively large storm event, reducing parameter estimation of other coefficients rather than only $k_1$ and $k_2$. 

The spillway formulation used in HydroPol2D is the same as HEC-RAS 2D, and the differences between both models are mainly because of the different reservoir stages due to the conceptualization of the 2D hydrodynamic solver and the culvert discharge that changes the stage. This difference, however, was almost negligible in this case, with NSE = 0.96 and KGE = 0.96, as shown in Fig.~\ref{fig:boundary}. The reduction in flow discharge of HydroPol2D in the spillway is proportional to the increase of discharge of HydroPol2D as shown in parts (b) and (a) of this figure, closing the mass balance at the outlet (c) with nearly no bias.

Rating curves as Eq.~\eqref{equ:boundary} were also used in other studies to underscore the importance of reservoir modeling with 2D models. The research conducted in \citep{Fleischmann2019-vt} used an adaptation of the MGB \citep{collischonn2007mgb} to simulate reservoirs as internal boundary conditions in the domain that used rating-curve equations as a replacement of the inertial momentum equation at the unit-catchment where the dam was located. As a result, the model allowed a more accurate representation of backwater and other hydrodynamics effects across the reservoir and its streams, favoring the proper estimation of the inundation extent and depth by allowing backwater modeling \citep{liro2022modelling}.

% This is literature review, not results and discussion
% In addition, research conducted in \citep{cea2022hydraulic} compared the use of empirical discharge equations with a 2D extension of the Two-component Pressure Approach (TPA) to simulate the observed data of bridges' backwater effect ; the comparing between the approaches determine that both can reproduce the observed data in a wide range of flow conditions, after a proper calibration and validation process. Other study that formulate an internal boundary condition in 2D models using as a reference an empirical discharge equation was \citep{morales2013formulation}; where author propose an equation to represents different conditions of connected gates of the main river with floodplains. In order to demonstrate its effectiveness in flood regulation and its potential for control through the modification of the internal boundary condition, it was implemented a Proportional-Derivative-Integral (PID) regulation algorithm.

\begin{figure}
    \centering
    \includegraphics[width=0.8\linewidth]{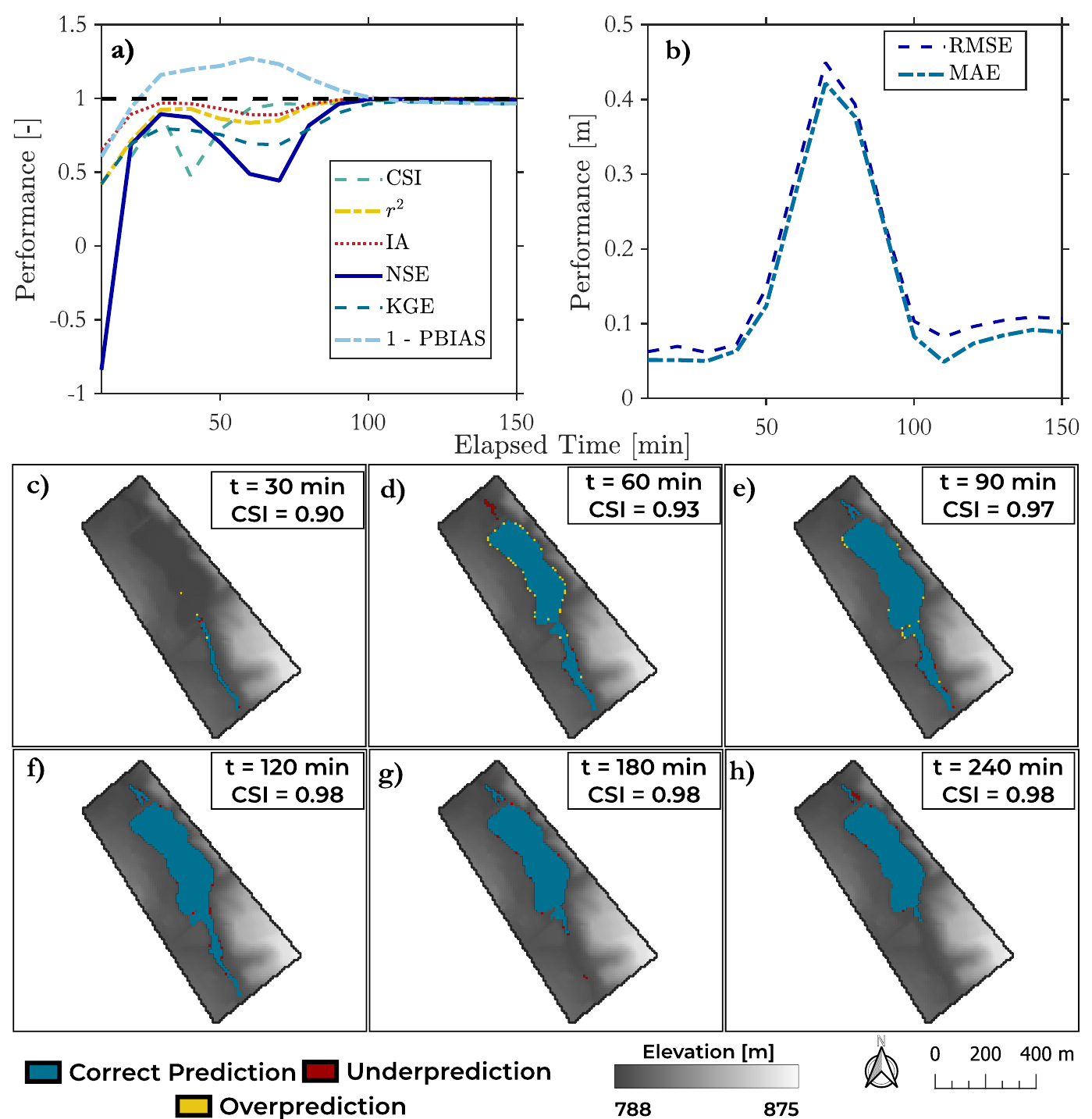}
    \caption{Spatio-temporal variation of model skillfulness comparing HydroPol2D (lim) with HEC-RAS 2D under a 100-yr inflow hydrograph in a detention pond with a culvert and a spillway flow controlling devices. Part (a) and (b) show the model performance using spatialized metrics, whereas (c) - (h) show the prediction performance for 30, 60, 90, 120, 180, and 240 minutes, respectively.}
    \label{fig:enter-label}
\end{figure}

%\begin{figure}
%    \centering
%    \includegraphics[width=1\linewidth]{Hydrograph_Constant.pdf}
%    \caption{Comparison between HydroPol2D and HEC-RAS discharge simulation under a steady-state inflow hydrograph of 50 $\mr{m^3 \cdot s^{-1}}$. Part (a) is the discharge at the culvert cell, (b) is at the spillway cell, and %(c) at the domain outlet.}
%    \label{fig:enter-label}
%\end{figure}

\subsection{Case Study No. 2}
The inclusion of internal conditions acting as stormwater structures significantly impacts the 2D water surface dynamics; this is expected since no inclusion of these conditions makes the inaccurate overland flow storage in few terrain areas due to noises and hydraulic discontinuities in the high-resolution terrain data, similarly as discussed in \citep{xing2022improving}. 

\begin{figure}
    \centering
    \includegraphics[scale = 0.75, trim={0.03cm 2cm 0.05cm 0.051cm},clip]{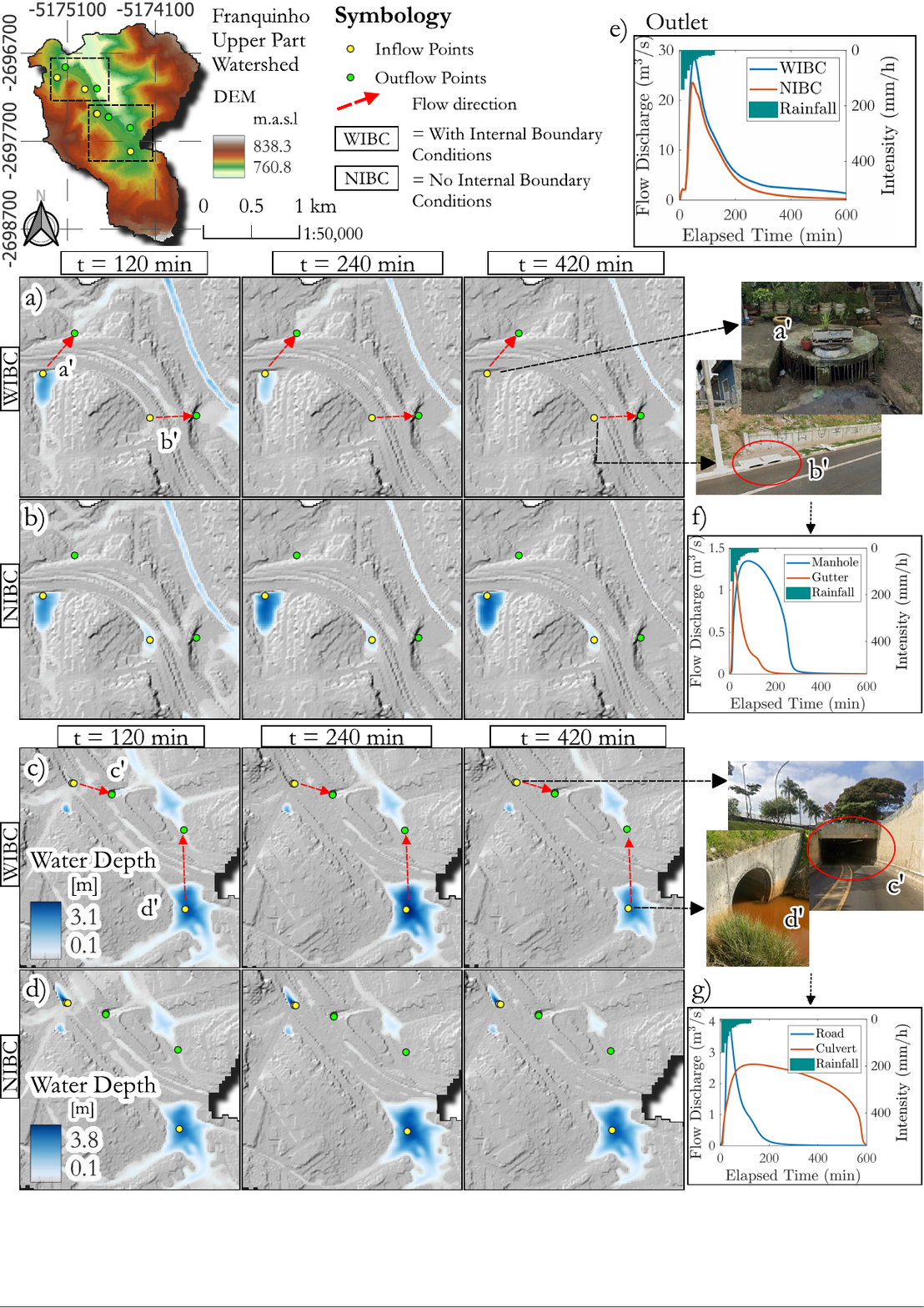}
    \caption{Comparison of simulated floodplain extent with the raw DEM with and without internal boundary conditions for the HydroPol2D (lim). Parts a) and b) are a manhole (a') and a curb cut (b') with and without internal boundary conditions, respectively; c) and d) are a road tunnel under a street (c') and a culvert (d') with and without internal boundary conditions, respectively; e) is the outlet hydrograph response between with and without internal boundary conditions; f) is the manhole and curb cut hydraulic behavior along the simulation; g) is the road tunnel under street and culvert hydraulic behavior along the simulation.}
    \label{fig:case 2 raw internal}
\end{figure}

There is a notable difference in floodplain extents and water depth magnitudes of approximately 1.8 meters throughout the simulation (Fig.~\ref{fig:case 2 raw internal}a) to Fig.~\ref{fig:case 2 raw internal}d), where the assessed structures were capable of draining the incoming overland flow. The smooth behavior of the urban drainage infrastructures can be seen in Fig.~\ref{fig:case 2 raw internal}e) and Fig.~\ref{fig:case 2 raw internal}f). It is important to note that water is stagnant in multiple minor areas, making the inclusion of these internal conditions necessary, as pointed out by \citep{towsif2020highly}.

% raw
% 3.54 m maximum depth
% 23.507 maximum outflow m3/s
% fix
% 1.66 m maximum depth
% 28.5 maximum outflow m3/s

Related to the outlet hydrograph comparisons between the two scenarios, Fig.~\ref{fig:case 2 raw internal}e) reveal differences in a peak discharge of 17.5\%, with a peak time difference of 5 minutes, and recession hydrograph shape difference, showing the continuous flow contribution from the assessed structures, reinforcing the fact of drainage structures inclusion in hydrodynamic modeling are crucial for accurate flood assessment in urban areas \citep{cui2019simulation}. Although the peak time difference was approximately 5 minutes at the outlet, at the boundary conditions, a completely different hydraulic behavior was seen with peak time differences of more than 60 minutes. Also, the assumptions used to represent drainage structure geometries effectively captured the rising and recession limb expected according to the design storm adopted (1\textsuperscript{st} Huff quartile, 2-h rainfall 50-yr rainfall). 

Nevertheless, the relation of the underestimated ($k_1$) parameter adopted for the culvert (d') showed in Fig.~\ref{fig:case 2 raw internal}d) and the larger drainage area into the culvert causes that point to remain flooded significantly longer than the other assessed points, and according to the Municipal Secretariat of Urban Infrastructure and Works from São Paulo \citep{SIURB}, there is no record of flooding from that area. Furthermore, notice that the road tunnel (c') in Fig.~\ref{fig:case 2 raw internal}c was simulated as a large culvert with high $k_1$ magnitude to avoid flow storage due to the terrain data quality. The latter highlights that a common calibration procedure it is necessary for the structures adopted for the modeling, and also enables a simplified alternative to overcome common issues when working with high-resolution terrain data \citep{abdullah2012improved}. The computational time, simulating with the internal boundary conditions, was approximately half of the case without urban drainage simulation inclusion.

% This increase in the execution time is due to the flow movement between neighbor pixels where flow is cumulating, on that conditions are provided more changes to the model become unstable if magnitudes of $\hat{h}$ are too low or near to flow depth tolerance, this will increase velocities, requiring lower temporal discretization to ensure model numerical stability, and consequently increasing the model time execution.

\subsection{Case Study No. 3}
This is a complex terrain in a coastal area with very mild slopes and diverse land use and land cover, increasing the difficulty level compared to previous case studies. In addition, a dam-break hydrograph with a peak flow of nearly 49,000 $\mr{m^3 \cdot s^{-1}}$ per 11 minutes is simulated, creating domain areas with Froude numbers above the unity. The spatio-temporal model skill for the dam-break simulation case study using the s-upwind scheme is shown in Fig.~\ref{fig:Temporal_Maps} (the interested reader can see this same figure for lim and s-centered in the supplemental information). This figure shows the temporal variation of the performance metrics and the spatio-temporal skillfulness of the model in comparison with HEC-RAS 2D full momentum solver. The temporal evaluation of the performance metrics is shown in Fig.~\ref{fig:Temporal_Maps} (a) and (b), where unitless metrics are shown in (a) and RMSE and MAE metrics are shown in (b). HydroPol2D consistently predicted larger flooded areas than HEC-RAS 2D for the lim, s-centered, and s-upwind schemes. Computational times, however, were at least 23 times faster for the slowest numerical scheme of HydroPol2D, which in this case was the s-upwind scheme. More specifically, HEC-RAS took approximately 2,610 min, whereas HydroPol2D with numerical schemes lim, s-centered, and s-upwind took approximately 99 min, 31 min, and 114 min, respectively. All computations were performed in an Intel(R) Core(TM) i7 10700F CPU @ 2.90GHz 2.90 GHz, 16GB RAM, NVIDIA GEFORCE GTX 1660 TI 6GB. RMSE and MAE metrics were relatively high at the beginning of the simulation and later resulted in values below 1 m after 60 min. 

For the s-upwind scheme, CSI values ranged from 0.46 to 0.92. The CSI for the maximum flood depth map was 0.89, indicating good performance \citep{bates2022flood}. The largest mismatches between HEC-RAS and HydroPol2D were due to the faster arrival of the flood wave in the city areas, which has very mild slopes (i.e.,  $ \leq 0.5^{\mr o}$ in most of the domain, see Fig.~\ref{fig:case_study_dam}). Water depths rapidly spread through the flat areas, increasing flooded areas and reducing wetted area model prediction, as shown in Fig.~\ref{fig:Temporal_Maps} for 20, 30, and 60 minutes after the dam-break. Overall, HydroPol2D can achieve better results in predicting wet areas for maximum flood depths (CSI = 0.95 [lim], 0.92 [s-centered], and 0.89 [s-upwind]). However, for areas closer to the dam breach, results can be affected by the lack of convective terms, especially shortly after the dam break. 

The model presented an excellent performance for predicting maximum benchmark correct flood areas, regardless of the numerical scheme, and could be easily adapted for simulating partial or total dam collapses in forecasting systems, especially because of the 20 times faster computational time. This would allow one to generate Monte-Carlo ensembles and probabilistic flood areas that could consider not only the model results but also the uncertainty in characteristics of the breach, initial reservoir storage, uncertainty in Manning's coefficients, and people exposure to hazards, as well as the conceptual model uncertainty.

\begin{figure}
    \centering
    \includegraphics[width=0.85\linewidth]{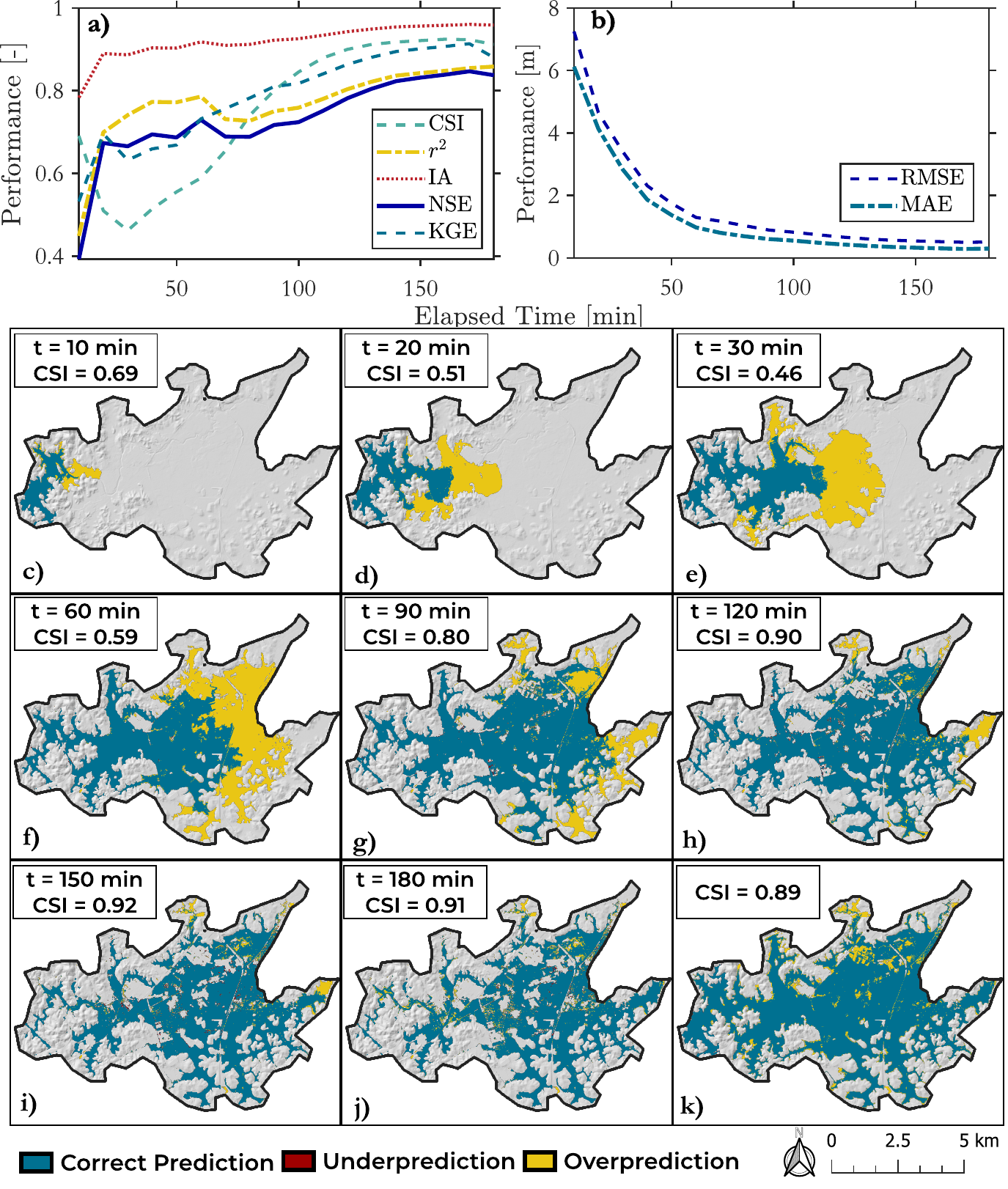}
    \caption{Time-varying performance of a dam-break scenario simulation comparing results of a local-inertial model with an upwind numerical scheme (HydroPol2D - upwind) with HEC-RAS 2D full dynamic model. Parts (a) and (b) show spatio-temporal metric indicators comparing HydroPol2D predictions with HEC-RAS predictions. The Black dashed line in part (a) indicates optimal performance. Parts (c) to (j) show the spatio-temporal prediction capacity of HydroPol2D for a threshold depth of 0.30 m for 10, 20, 30, 60, 90, 120, 150, and 180 minutes after the dam break. Part (k) shows the prediction capacity for the maximum flood depth maps.}
    \label{fig:Temporal_Maps}
\end{figure}

\begin{figure}
    \centering
    \includegraphics[width=1\linewidth]{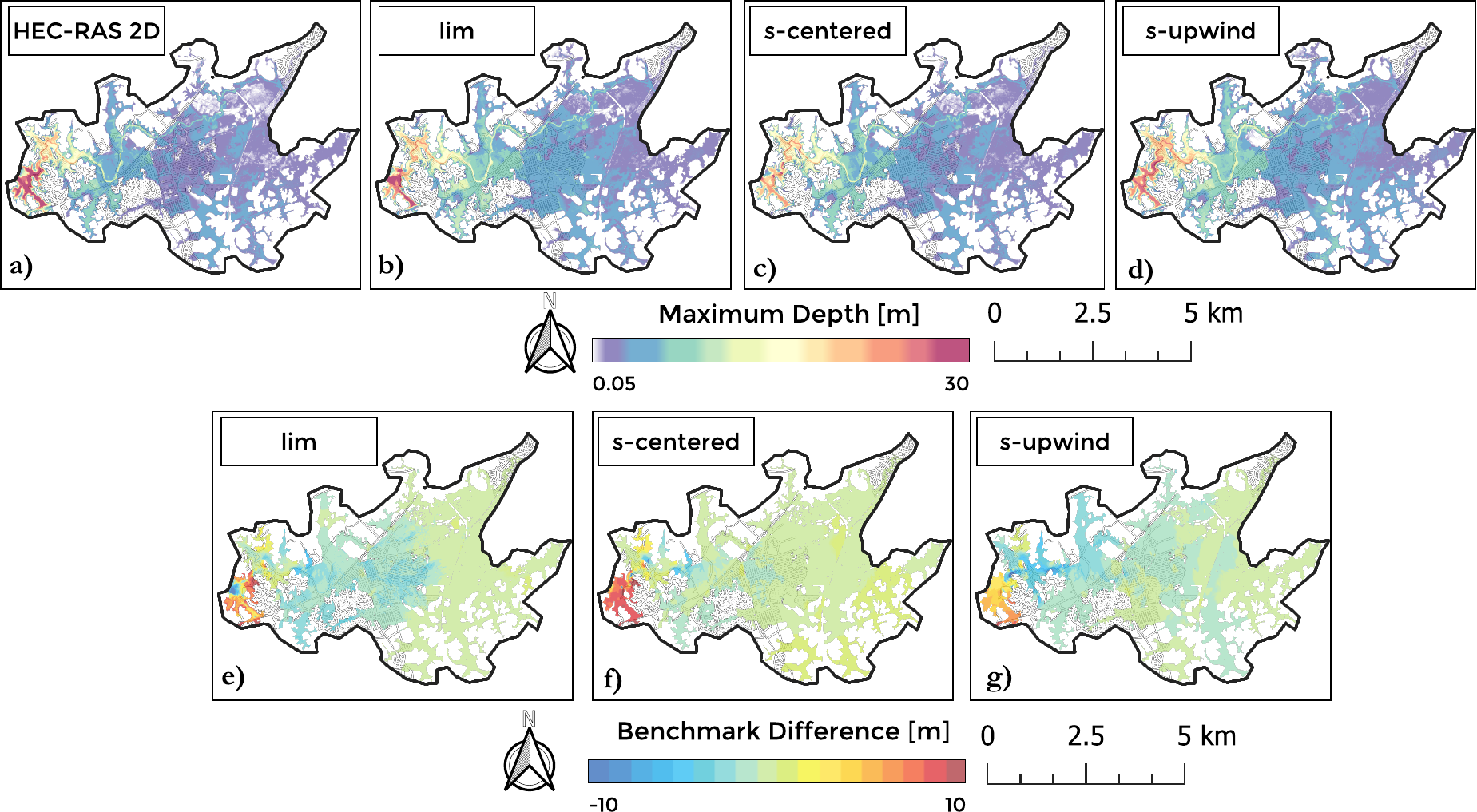}
    \caption{Maximum flood depth for (a) HEC-RAS 2D, (b) local-inertial model, (c) s-centered scheme, and (d) s-upwind scheme. Parts (e) to (g) shows the difference between HEC-RAS 2D full momentum solver against the local inertial original formulation (e), s-centered scheme (f), and (g) s-upwind scheme, where positive values indicate that HEC-RAS predicted larger depths and negative values indicate that HydroPol2D predicted larger depths, comparing both models.}
    \label{fig:Maximum_Depths}
\end{figure}

A comparison of maximum flood depths against the benchmark model presented in Fig.~\ref{fig:Maximum_Depths} shows a different scenario. The flood prediction at the city level was predicted with relatively good accuracy; however, larger differences arise close to the dam. As expected, these areas of potential high froud number were the ones that caused larger discrepancies between a full momentum solver and a local-inertial model. In particular, the largest differences occurred for the s-centered scheme at the region close to the inflow hydrograph; however, interestingly, the s-centered scheme had the largest CSI value for the maximum flood depth (see SI where CSI values over time are plotted for lim and s-centered). The scheme with acceptable CSI performance and better NSE, KGE, and RMSE metrics is the s-upwind scheme, comparing all methods. This indicates that this scheme can be more applicable to accurately predicting wet areas and estimating maximum flood depths.

\begin{figure}
    \centering
    \includegraphics[width=1\linewidth]{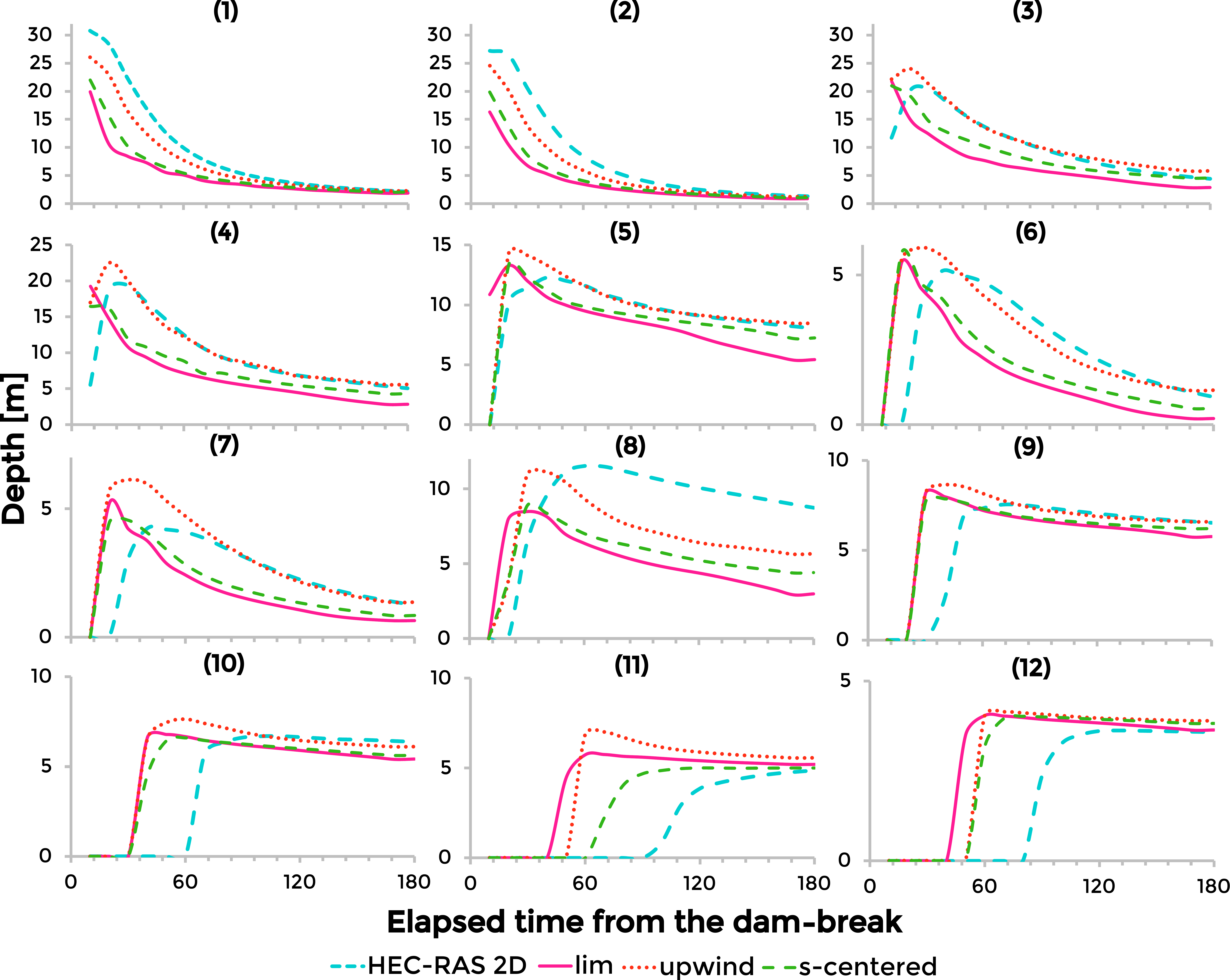}
    \caption{Simulation results of a dam-break using different local-inertial conceptualizations compared to a 2D full momentum benchmark (HEC-RAS 2D) for the waterlogging points shown in Fig.~\ref{fig:case_study_dam}.}    \label{fig:Waterlog_Points}
\end{figure}

The temporal evolution of the flood depths in the waterlogging points of Fig.~\ref{fig:case_study_dam} is shown in Fig.~\ref{fig:Waterlog_Points}. Points (1) - (5) are located in the river path, Points (6) - (7) are located in an urbanized area of the city, point (11) is located in a tributary of the main river, whereas points 8, 9, 10, and 12 are following the main river path. This figure shows that the s-upwind had the largest agreement with HEC-RAS, especially for points closer to the dam. Maximum depths are relatively predicted correct for the farthest points (9-12), at the cost of inaccurate arrival times. The model presented biased results for the urban area flood depths (7-8). Tab.~\ref{tab:performance_points} shows the typical hydrological performance metrics to assess model fitness for the waterlogging points. The model performance for all schemes was fairly good for points 1-5, with a reducing performance as the flood wave progresses toward the flat domain in the city's urban area and later towards the outlet.

\begin{table}
\centering
\begin{tblr}{
  hline{1-2,14} = {-}{},
}
{Point} & $r^2$[-]       & IA~[-]             & NSE~[-]              & KGE [-]             & RMSE [m]           \\
1                    & (0.87, \textbf{0.99}, 0.93) & (0.77, \textbf{0.97} , 0.86) & (0.43, \textbf{0.91}, 0.62)   & (0.30, \textbf{0.72}, 0.43)   & (6.79, \textbf{2.65}, 5.53) \\
2                    & (0.90, \textbf{0.97}, 0.91)  & (0.76, \textbf{0.96}, 0.85) & (0.40, \textbf{0.88}, 0.59)    & (0.24, \textbf{0.68}, 0.39)  & (6.53, \textbf{2.86}, 5.35) \\
3                    & (0.49, \textbf{0.84}, 0.65) & (0.75, \textbf{0.94,} 0.87) & (0.03, \textbf{0.72}, 0.55)   & (0.56, \textbf{0.79}, 0.76)  & (5.05, \textbf{2.72}, 3.43) \\
4                    & (0.26, \textbf{0.73}, 0.42) & (0.66, \textbf{0.92}, 0.76) & (-0.28, \textbf{0.63}, 0.25)  & (0.42, \textbf{0.80}, 0.55)   & (5.29, \textbf{2.86}, 4.07) \\
5                    & (0.03, \textbf{0.88}, 0.84) & (0.42, \textbf{0.95}, 0.94) & (-0.56, \textbf{0.78}, 0.78)  & (0.15, 0.81, \textbf{0.89})  & (3.26, \textbf{1.21}, 1.23) \\
6                    & (0.17, \textbf{0.45}, 0.20)  & (0.63, \textbf{0.83}, 0.68) & (-0.49, \textbf{0.26}, -0.21) & (0.29, \textbf{0.65}, 0.39)  & (2.00, \textbf{1.41}, 1.81)    \\
7                    & (0.09, \textbf{0.36}, 0.21) & (0.59, \textbf{0.74}, 0.69) & (-0.75, -0.57, \textbf{-0.2}) & (0.25, 0.37, \textbf{0.43})  & (1.70, 1.61, \textbf{1.41})  \\
8                    & (0.02, \textbf{0.42}, 0.37) & (0.41, \textbf{0.74}, 0.60)  & (-1.55, \textbf{0.08}, -0.62) & (-0.02, \textbf{0.53}, 0.31) & (5.33, \textbf{3.21}, 4.24) \\
9                    & (0.30, 0.36, \textbf{0.37})  & (0.75, \textbf{0.78}, 0.78) & (0.22, 0.18, \textbf{0.31})   & (0.50, \textbf{0.55}, 0.55)   & (2.43, 2.48, \textbf{2.29}) \\
10                   & (0.27, 0.29, \textbf{0.40})  & (0.73, 0.74, \textbf{0.78}) & (0.16, 0.04, \textbf{0.34})   & (0.43, 0.43, \textbf{0.52})  & (2.80, 2.99, \textbf{2.48})  \\
11                   & (0.21, 0.20, \textbf{0.58)}  & (0.63, 0.61, \textbf{0.83)} & (-1.24, -1.82, \textbf{0.19}) & (-0.26, -0.40, \textbf{0.36)} & (3.19, 3.58, \textbf{1.92}) \\
12                   & (0.31, 0.44, \textbf{0.50})  & (0.73, 0.78, \textbf{0.81}) & (-0.18, -0.02, \textbf{0.15}) & (0.29, 0.39, \textbf{0.45})  & (1.88, 1.75, \textbf{1.6})  
\end{tblr}
\caption{Performance indicators at waterlogging points. The entries in this table follow (lim, s-upwind, s-centered) order. The bold values represent the most suitable metric within the three methods tested.}
\label{tab:performance_points}
\end{table}

% Floodplain extents for 30 m resolution and 5 m resolution compared to HEC-RAS. Calculate Hit, Misses, and False Alarms.

\section{Conclusions}
Applying low-complexity models has been extensively tested in the literature for typical flood hydrological conditions where the convective acceleration can be safely neglected. In this paper, we provide 4 testing cases where this acceleration can play an important role. We investigate not only the performance of the model to predict maximum flood extents and depths but also the spatio-temporal performance of the model compared to a benchmark fully hydrodynamic model. Three numerical schemes were tested, named the original local inertial formulation (lim), (s-centered), and (s-upwind). The results of this paper support the following conclusions:

\begin{itemize}
    \item Imposing internal boundary conditions in the domain to represent the detention pond culvert and spillway presented great performance (NSE and KGE $>$ 0.96) when compared to a fully hydrodynamic model that solves depth-varying culvert equations and spillway equations for a 1 in 100-year inflow hydrograph. The modeling approach developed to simulate the boundary conditions only requires 2 parameters that can be derived from local inspections, design plans, or ultimately via freely available GIS data. For culverts, these represent the culvert area and the power law culvert coefficient.
    \item Results of the application for a 1 in 50 years, 2 hours Huff distributed rainfall in an urbanized catchment with LiDAR data incorrectly representing the urban features as culverts and tunnels show an incorrect representation of the hydraulic behavior of the city infrastructure. The lack of internal boundary conditions resulted in a peak discharge difference of nearly 17.5\% at the outlet. Locally at the internal boundary conditions, incorrect hydrodynamic behaviors are seen with flows only leaving cells if water depth rises above the LiDAR, which incorrectly would represent the case of the manholes, inlets, and culverts simulated in the catchment. This results in flood depths of 1.8 m higher than when simulating with the internal boundary conditions representing the urban drainage infrastructure, resulting in approximately two times larger computational times.
    \item The spatio-temporal performance of the different local inertial numerical schemes for a dam-break, resulting in an inflow hydrograph of approximately 49,000 $\mr{m^3 \cdot s^{-1}}$ was investigated. The CSI metrics for the maximum flood depths for the lim, s-centered, and s-upwind were 0.92, 0.89, and 0.95, respectively. Although a larger CSI was found for s-centered and lim, the s-upwind had superior performance metrics of NSE, KGE, RMSE, $r^2$, and IA for the spatio-temporal analysis, indicating that this scheme is better suitable to predict not only accurate maximum flood extents but also flood depths depending on the level of accuracy required for the application. Overall, an advance of the flood wave extent was observed for all local inertial schemes tested due to the lack of advective inertial. These results suggest that local-inertial model applications focused only on determining flood inundation extents are more accurate than those predicting arrival times and inundation depths.    
\end{itemize}
Future studies can apply the modeling framework to ensemble and probabilistic analysis of dam breaks and also coupling data assimilation techniques such as particle filters of ensemble Kalman filters to attach the lack of conceptual capability with observed data, locally correcting states and parameters, especially under highly-variable flows.

\section{Supplemental Information}
The supplemental information describes inflow hydrographs, $k_1$ and $k_2$, and parameters used for case study 1. In addition, it details the internal boundary conditions parametrization used in case study 2 and shows the hyetograph used. Finally, it shows the inflow hydrograph, slope map, CSI metrics for different numerical schemes, and HEC-RAS and HydroPol2D temporal depths for case study 3.

\section{Apendix - Performance Indicators}
The raster-based analysis presented in this paper is performed only when both models have data. Raster data is concatenated, and performance metrics are calculated using the vectorized raster values.

\subsection{Nash-Sutcliffe-Efficiency} \label{sec:NSE}
The Nash-Sutcliffe-Efficiency (NSE) metric is calculated in terms of the observed variable (e.g., generally flow discharge) and the modeled variable such that:

\begin{equation} \label{equ:nash}
\mathrm{NSE}=1-\frac{\sum_{i=1}^{n_{\mr{obs}}}\left(y_{\mathrm{obs}}^i-y_{\mathrm{m}}^i\right)^2}{\sum_{i=1}^{n_{\mr{obs}}}\left(y^i_{\mathrm{obs}}-\overline{y_{\mathrm{obs}}}\right)^2}
\end{equation}
where $y_{\mathrm{obs}}$ is the observed or the assumed true variable, whereas $y_{\mathrm m}$ is the variable. The indexes herein expressed as $i$ and $n_{\mr{obs}}$ represent the time in which the observations were made and the number of observations, respectively. NSE ranges from $- \infty$ to 1 (inclusive), with negative values indicating that the observed mean has smaller squared error than the modeled results. Ideally, a NSE = 1 indicates a perfect match between modeled and observed values.

\subsection{Coefficient of Determination} \label{sec:r2}
The coefficient of determination determines the correlation between the observations. It ranges from 0 to 1, with 1 corresponding to a perfect correlation between the modeled and observed data, and can be calculated as:

\begin{equation}
r^2=\left(\frac{\sum_{i=1}^{n_{\mr{obs}}}\left(y_{\mathrm{m}}-\overline{y_{\mathrm{m}}^i}\right)\left(y_{\mathrm{obs}}^i-\overline{y_{\mathrm{obs}}^i}\right)}{\sqrt{\sum_{i=1}^{n_{\mr{obs}}}\left(y_{\mathrm{m}}-\overline{y_{\mathrm{m}}^i}\right)^2 \sum_{i=1}^i\left(y_{\mathrm{obs}}^i-\overline{y_{\mathrm{obs}}}\right)^2}}\right)^2
\end{equation}

\subsection{KGE}
The KGE (Kling-Gupta Efficiency) is a measure of model performance that takes into account three components: linear correlation between observed and simulated records (\textit{cc}), variability in the data values ($\alpha$, equal to the standard deviation of simulated over the standard deviation observed), and bias ($\beta$, equal to the mean of simulated over the mean of observed). For a perfect model with no data errors, the value of KGE is 1. The KGE metric is written as:

\begin{equation}
\mathrm{KGE} =1-\sqrt{(c c-1)^2+(\alpha-1)^2+(\beta-1)^2}
\end{equation}
\subsection{Root-Mean-Square-Error} \label{sec:RMSE}
The Root-Mean-Square-Error (RMSE) index measures the average difference between predicted and observed variables and can be calculated as follows:

\begin{equation}
\mathrm{RMSE}=\sqrt{\frac{\sum_{i=1}^{n_{\mr{obs}}}\left(y_{m }^i-y_{\mathrm{obs}}^i\right)^2}{n}}
\end{equation}

\subsection{Mean Average Error}
The mean average error calculates mean error between the benchmark and the model and is given by:

\begin{equation}
    \mathrm{MAE}=\frac{1}{n} \sum_{i=1}^n \abs{y_{\mathrm{obs}}^i-y_{\mathrm{m}}^i}
\end{equation}

\subsection{Critical Sucess Index}

To measure the model capability to predict flood depths, the Critical Sucess Index (CSI) metric to compare maps of flood depths between HydroPol2D and HEC-RAS 2D full momentum solver. Typically, CSI analysis are performed only for maximum flood extents; however, in this paper, we assess the temporal variation of these metrics for different times. The CSI metric measures the model probability to match with the proxy observations based on the summation of pixels that attend a certain threshold, and is written as:

\begin{equation}
\text { CSI }=\frac{\text { Correct Predictions }}{\text { Correct Predictions }+ \text { Overpredictions }+ \text { Underpredictions }}
\end{equation}
where the Correct Predictions sum all cells that exceed a certain depth $\delta$, herein assumed as 0.30 m \citep{do2023generalizing}. The overprediction is the total of cells that only HydroPol2D predicted flooded and underpredictions are the total that only HEC-RAS predicted.

\subsection{Index of Agreement}
The index of agreement (IA) is calculated by dividing the mean squared error (MSE) of the prediction by the variance of the observed data. Values closer to unity indicate better model performance. The IA is given by:

\begin{equation}
I A=1-\frac{\sum_{i=1}^n\left(y^i_{\mr{obs}}-y^i_{\mr{m}}\right)^2}{\sum_{i=1}^n\left(\left|y^i_{\mr{m}}-\overline{y_{\mathrm{obs}}}\right|+\left|y^i_{\mr{obs}}-\overline{y_{\mathrm{obs}}}\right|\right)^2}
\end{equation}

\section{Acknowledgments} ~\label{sec:acknow}

We would like to acknowledge the Husqvarna group for sharing land use and land cover data for the study cases in the São Paulo city, Brazil.

\bibliographystyle{cas-model2-names}

\bibliography{References}

\begin{thebibliography}{72}
\expandafter\ifx\csname natexlab\endcsname\relax\def\natexlab#1{#1}\fi
\providecommand{\url}[1]{\texttt{#1}}
\providecommand{\href}[2]{#2}
\providecommand{\path}[1]{#1}
\providecommand{\DOIprefix}{doi:}
\providecommand{\ArXivprefix}{arXiv:}
\providecommand{\URLprefix}{URL: }
\providecommand{\Pubmedprefix}{pmid:}
\providecommand{\doi}[1]{\href{http://dx.doi.org/#1}{\path{#1}}}
\providecommand{\Pubmed}[1]{\href{pmid:#1}{\path{#1}}}
\providecommand{\bibinfo}[2]{#2}
\ifx\xfnm\relax \def\xfnm[#1]{\unskip,\space#1}\fi
%Type = Article
\bibitem[{Abdullah et~al.(2012)Abdullah, Vojinovic, Price and Aziz}]{abdullah2012improved}
\bibinfo{author}{Abdullah, A.}, \bibinfo{author}{Vojinovic, Z.}, \bibinfo{author}{Price, R.}, \bibinfo{author}{Aziz, N.}, \bibinfo{year}{2012}.
\newblock \bibinfo{title}{Improved methodology for processing raw lidar data to support urban flood modelling--accounting for elevated roads and bridges}.
\newblock \bibinfo{journal}{Journal of Hydroinformatics} \bibinfo{volume}{14}, \bibinfo{pages}{253--269}.
%Type = Article
\bibitem[{Afshari et~al.(2018)Afshari, Tavakoly, Rajib, Zheng, Follum, Omranian and Fekete}]{afshari2018comparison}
\bibinfo{author}{Afshari, S.}, \bibinfo{author}{Tavakoly, A.A.}, \bibinfo{author}{Rajib, M.A.}, \bibinfo{author}{Zheng, X.}, \bibinfo{author}{Follum, M.L.}, \bibinfo{author}{Omranian, E.}, \bibinfo{author}{Fekete, B.M.}, \bibinfo{year}{2018}.
\newblock \bibinfo{title}{Comparison of new generation low-complexity flood inundation mapping tools with a hydrodynamic model}.
\newblock \bibinfo{journal}{Journal of Hydrology} \bibinfo{volume}{556}, \bibinfo{pages}{539--556}.
%Type = Article
\bibitem[{Aronica et~al.(2012)Aronica, Franza, Bates and Neal}]{aronica2012probabilistic}
\bibinfo{author}{Aronica, G.T.}, \bibinfo{author}{Franza, F.}, \bibinfo{author}{Bates, P.}, \bibinfo{author}{Neal, J.}, \bibinfo{year}{2012}.
\newblock \bibinfo{title}{Probabilistic evaluation of flood hazard in urban areas using monte carlo simulation}.
\newblock \bibinfo{journal}{Hydrological Processes} \bibinfo{volume}{26}, \bibinfo{pages}{3962--3972}.
%Type = Article
\bibitem[{Bates(2023)}]{bates2023fundamental}
\bibinfo{author}{Bates, P.}, \bibinfo{year}{2023}.
\newblock \bibinfo{title}{Fundamental limits to flood inundation modelling}.
\newblock \bibinfo{journal}{Nature Water} \bibinfo{volume}{1}, \bibinfo{pages}{566--567}.
%Type = Article
\bibitem[{Bates(2022)}]{bates2022flood}
\bibinfo{author}{Bates, P.D.}, \bibinfo{year}{2022}.
\newblock \bibinfo{title}{Flood inundation prediction}.
\newblock \bibinfo{journal}{Annual Review of Fluid Mechanics} \bibinfo{volume}{54}, \bibinfo{pages}{287--315}.
%Type = Article
\bibitem[{Bates et~al.(2010)Bates, Horritt and Fewtrell}]{bates2010simple}
\bibinfo{author}{Bates, P.D.}, \bibinfo{author}{Horritt, M.S.}, \bibinfo{author}{Fewtrell, T.J.}, \bibinfo{year}{2010}.
\newblock \bibinfo{title}{A simple inertial formulation of the shallow water equations for efficient two-dimensional flood inundation modelling}.
\newblock \bibinfo{journal}{Journal of hydrology} \bibinfo{volume}{387}, \bibinfo{pages}{33--45}.
%Type = Article
\bibitem[{Bellos and Tsakiris(2015)}]{bellos2015comparing}
\bibinfo{author}{Bellos, V.}, \bibinfo{author}{Tsakiris, G.}, \bibinfo{year}{2015}.
\newblock \bibinfo{title}{Comparing various methods of building representation for 2d flood modelling in built-up areas}.
\newblock \bibinfo{journal}{Water Resources Management} \bibinfo{volume}{29}, \bibinfo{pages}{379--397}.
%Type = Article
\bibitem[{Brown et~al.(2022)Brown, Brumby, Guzder-Williams, Birch, Hyde, Mazzariello, Czerwinski, Pasquarella, Haertel, Ilyushchenko et~al.}]{brown2022dynamic}
\bibinfo{author}{Brown, C.F.}, \bibinfo{author}{Brumby, S.P.}, \bibinfo{author}{Guzder-Williams, B.}, \bibinfo{author}{Birch, T.}, \bibinfo{author}{Hyde, S.B.}, \bibinfo{author}{Mazzariello, J.}, \bibinfo{author}{Czerwinski, W.}, \bibinfo{author}{Pasquarella, V.J.}, \bibinfo{author}{Haertel, R.}, \bibinfo{author}{Ilyushchenko, S.}, et~al., \bibinfo{year}{2022}.
\newblock \bibinfo{title}{Dynamic world, near real-time global 10 m land use land cover mapping}.
\newblock \bibinfo{journal}{Scientific Data} \bibinfo{volume}{9}, \bibinfo{pages}{251}.
%Type = Article
\bibitem[{Brunner(2016)}]{Brunner}
\bibinfo{author}{Brunner, G.}, \bibinfo{year}{2016}.
\newblock \bibinfo{title}{Hec-ras river analysis system, 2d modeling user’s manual, version 5.0}.
\newblock \bibinfo{journal}{Davis: US Army Corps of Engineers, hydrologic engineering center} .
%Type = Article
\bibitem[{Camporese et~al.(2010)Camporese, Paniconi, Putti and Orlandini}]{camporese2010surface}
\bibinfo{author}{Camporese, M.}, \bibinfo{author}{Paniconi, C.}, \bibinfo{author}{Putti, M.}, \bibinfo{author}{Orlandini, S.}, \bibinfo{year}{2010}.
\newblock \bibinfo{title}{Surface-subsurface flow modeling with path-based runoff routing, boundary condition-based coupling, and assimilation of multisource observation data}.
\newblock \bibinfo{journal}{Water Resources Research} \bibinfo{volume}{46}.
%Type = Book
\bibitem[{Canholi(2015)}]{canholi2015drenagem}
\bibinfo{author}{Canholi, A.}, \bibinfo{year}{2015}.
\newblock \bibinfo{title}{Drenagem urbana e controle de enchentes}.
\newblock \bibinfo{publisher}{Oficina de textos}.
%Type = Article
\bibitem[{Chang et~al.(2015)Chang, Wang and Chen}]{chang2015novel}
\bibinfo{author}{Chang, T.J.}, \bibinfo{author}{Wang, C.H.}, \bibinfo{author}{Chen, A.S.}, \bibinfo{year}{2015}.
\newblock \bibinfo{title}{A novel approach to model dynamic flow interactions between storm sewer system and overland surface for different land covers in urban areas}.
\newblock \bibinfo{journal}{Journal of Hydrology} \bibinfo{volume}{524}, \bibinfo{pages}{662--679}.
%Type = Article
\bibitem[{Chen et~al.(2018)Chen, Liang, Liu and Xie}]{chen2018hydraulic}
\bibinfo{author}{Chen, H.}, \bibinfo{author}{Liang, Q.}, \bibinfo{author}{Liu, Y.}, \bibinfo{author}{Xie, S.}, \bibinfo{year}{2018}.
\newblock \bibinfo{title}{Hydraulic correction method (hcm) to enhance the efficiency of srtm dem in flood modeling}.
\newblock \bibinfo{journal}{Journal of hydrology} \bibinfo{volume}{559}, \bibinfo{pages}{56--70}.
%Type = Article
\bibitem[{Collischonn et~al.(2007)Collischonn, Allasia, Da~Silva and Tucci}]{collischonn2007mgb}
\bibinfo{author}{Collischonn, W.}, \bibinfo{author}{Allasia, D.}, \bibinfo{author}{Da~Silva, B.C.}, \bibinfo{author}{Tucci, C.E.}, \bibinfo{year}{2007}.
\newblock \bibinfo{title}{The mgb-iph model for large-scale rainfall—runoff modelling}.
\newblock \bibinfo{journal}{Hydrological Sciences Journal} \bibinfo{volume}{52}, \bibinfo{pages}{878--895}.
%Type = Article
\bibitem[{Cui et~al.(2019)Cui, Liang, Wang, Zhao, Hu, Wang and Xia}]{cui2019simulation}
\bibinfo{author}{Cui, Y.}, \bibinfo{author}{Liang, Q.}, \bibinfo{author}{Wang, G.}, \bibinfo{author}{Zhao, J.}, \bibinfo{author}{Hu, J.}, \bibinfo{author}{Wang, Y.}, \bibinfo{author}{Xia, X.}, \bibinfo{year}{2019}.
\newblock \bibinfo{title}{Simulation of hydraulic structures in 2d high-resolution urban flood modeling}.
\newblock \bibinfo{journal}{Water} \bibinfo{volume}{11}, \bibinfo{pages}{2139}.
%Type = Article
\bibitem[{Dazzi et~al.(2018)Dazzi, Vacondio, Dal~Pal{\`u} and Mignosa}]{dazzi2018local}
\bibinfo{author}{Dazzi, S.}, \bibinfo{author}{Vacondio, R.}, \bibinfo{author}{Dal~Pal{\`u}, A.}, \bibinfo{author}{Mignosa, P.}, \bibinfo{year}{2018}.
\newblock \bibinfo{title}{A local time stepping algorithm for gpu-accelerated 2d shallow water models}.
\newblock \bibinfo{journal}{Advances in water resources} \bibinfo{volume}{111}, \bibinfo{pages}{274--288}.
%Type = Article
\bibitem[{De~Almeida and Bates(2013)}]{de2013applicability}
\bibinfo{author}{De~Almeida, G.A.}, \bibinfo{author}{Bates, P.}, \bibinfo{year}{2013}.
\newblock \bibinfo{title}{Applicability of the local inertial approximation of the shallow water equations to flood modeling}.
\newblock \bibinfo{journal}{Water Resources Research} \bibinfo{volume}{49}, \bibinfo{pages}{4833--4844}.
%Type = Article
\bibitem[{De~Roo et~al.(2000)De~Roo, Wesseling and Van~Deursen}]{de2000physically}
\bibinfo{author}{De~Roo, A.}, \bibinfo{author}{Wesseling, C.}, \bibinfo{author}{Van~Deursen, W.}, \bibinfo{year}{2000}.
\newblock \bibinfo{title}{Physically based river basin modelling within a gis: the lisflood model}.
\newblock \bibinfo{journal}{Hydrological Processes} \bibinfo{volume}{14}, \bibinfo{pages}{1981--1992}.
%Type = Article
\bibitem[{Downer and Ogden(2004)}]{downer2004gssha}
\bibinfo{author}{Downer, C.W.}, \bibinfo{author}{Ogden, F.L.}, \bibinfo{year}{2004}.
\newblock \bibinfo{title}{Gssha: Model to simulate diverse stream flow producing processes}.
\newblock \bibinfo{journal}{Journal of Hydrologic Engineering} \bibinfo{volume}{9}, \bibinfo{pages}{161--174}.
%Type = Article
\bibitem[{Dung et~al.(2011)Dung, Merz, B{\'a}rdossy, Thang and Apel}]{dung2011multi}
\bibinfo{author}{Dung, N.V.}, \bibinfo{author}{Merz, B.}, \bibinfo{author}{B{\'a}rdossy, A.}, \bibinfo{author}{Thang, T.D.}, \bibinfo{author}{Apel, H.}, \bibinfo{year}{2011}.
\newblock \bibinfo{title}{Multi-objective automatic calibration of hydrodynamic models utilizing inundation maps and gauge data}.
\newblock \bibinfo{journal}{Hydrology and Earth System Sciences} \bibinfo{volume}{15}, \bibinfo{pages}{1339--1354}.
%Type = Article
\bibitem[{Fleischmann et~al.(2019a)Fleischmann, Collischonn, Paiva and Tucci}]{fleischmann2019modeling}
\bibinfo{author}{Fleischmann, A.}, \bibinfo{author}{Collischonn, W.}, \bibinfo{author}{Paiva, R.}, \bibinfo{author}{Tucci, C.E.}, \bibinfo{year}{2019}a.
\newblock \bibinfo{title}{Modeling the role of reservoirs versus floodplains on large-scale river hydrodynamics}.
\newblock \bibinfo{journal}{Natural Hazards} \bibinfo{volume}{99}, \bibinfo{pages}{1075--1104}.
%Type = Article
\bibitem[{Fleischmann et~al.(2019b)Fleischmann, Collischonn, Paiva and Tucci}]{Fleischmann2019-vt}
\bibinfo{author}{Fleischmann, A.}, \bibinfo{author}{Collischonn, W.}, \bibinfo{author}{Paiva, R.}, \bibinfo{author}{Tucci, C.E.}, \bibinfo{year}{2019}b.
\newblock \bibinfo{title}{Modeling the role of reservoirs versus floodplains on large-scale river hydrodynamics}.
\newblock \bibinfo{journal}{Nat. Hazards (Dordr.)} \bibinfo{volume}{99}, \bibinfo{pages}{1075--1104}.
%Type = Article
\bibitem[{Fleischmann et~al.(2020)Fleischmann, Paiva, Collischonn, Siqueira, Paris, Moreira, Papa, Bitar, Parrens, Aires et~al.}]{fleischmann2020trade}
\bibinfo{author}{Fleischmann, A.}, \bibinfo{author}{Paiva, R.}, \bibinfo{author}{Collischonn, W.}, \bibinfo{author}{Siqueira, V.}, \bibinfo{author}{Paris, A.}, \bibinfo{author}{Moreira, D.}, \bibinfo{author}{Papa, F.}, \bibinfo{author}{Bitar, A.}, \bibinfo{author}{Parrens, M.}, \bibinfo{author}{Aires, F.}, et~al., \bibinfo{year}{2020}.
\newblock \bibinfo{title}{Trade-offs between 1-d and 2-d regional river hydrodynamic models}.
\newblock \bibinfo{journal}{Water Resources Research} \bibinfo{volume}{56}, \bibinfo{pages}{e2019WR026812}.
%Type = Article
\bibitem[{George and Nair(2015)}]{george2015dam}
\bibinfo{author}{George, A.C.}, \bibinfo{author}{Nair, B.T.}, \bibinfo{year}{2015}.
\newblock \bibinfo{title}{Dam break analysis using boss dambrk}.
\newblock \bibinfo{journal}{Aquatic Procedia} \bibinfo{volume}{4}, \bibinfo{pages}{853--860}.
%Type = Article
\bibitem[{Getirana et~al.(2023)Getirana, Mandarino, de~Montezuma and Kirschbaum}]{getirana2023urban}
\bibinfo{author}{Getirana, A.}, \bibinfo{author}{Mandarino, F.}, \bibinfo{author}{de~Montezuma, P.N.}, \bibinfo{author}{Kirschbaum, D.}, \bibinfo{year}{2023}.
\newblock \bibinfo{title}{An urban drainage scheme for large-scale flood models}.
\newblock \bibinfo{journal}{Journal of Hydrology} \bibinfo{volume}{627}, \bibinfo{pages}{130410}.
%Type = Article
\bibitem[{Getirana and Paiva(2013)}]{getirana2013mapping}
\bibinfo{author}{Getirana, A.C.}, \bibinfo{author}{Paiva, R.C.}, \bibinfo{year}{2013}.
\newblock \bibinfo{title}{Mapping large-scale river flow hydraulics in the amazon basin}.
\newblock \bibinfo{journal}{Water Resources Research} \bibinfo{volume}{49}, \bibinfo{pages}{2437--2445}.
%Type = Article
\bibitem[{Gomes~Jr et~al.(2024a)Gomes~Jr, Castro, da~Silva, Giacomoni and Mendiondo}]{gomes2024increasing}
\bibinfo{author}{Gomes~Jr, M.N.}, \bibinfo{author}{Castro, M.d.A.R.A.}, \bibinfo{author}{da~Silva, P.G.C.}, \bibinfo{author}{Giacomoni, M.H.}, \bibinfo{author}{Mendiondo, E.M.}, \bibinfo{year}{2024}a.
\newblock \bibinfo{title}{Increasing flood awareness through dam-break serious games}.
\newblock \bibinfo{journal}{International Journal of Disaster Risk Reduction} , \bibinfo{pages}{104543}.
%Type = Article
\bibitem[{Gomes~Jr et~al.(2024b)Gomes~Jr, Giacomoni, Navarro and Mendiondo}]{gomes2024global}
\bibinfo{author}{Gomes~Jr, M.N.}, \bibinfo{author}{Giacomoni, M.H.}, \bibinfo{author}{Navarro, F.A.R.}, \bibinfo{author}{Mendiondo, E.M.}, \bibinfo{year}{2024}b.
\newblock \bibinfo{title}{Global optimization-based calibration algorithm for a 2d distributed hydrologic-hydrodynamic and water quality model}.
\newblock \bibinfo{journal}{Environmental Modelling \& Software} , \bibinfo{pages}{106128}.
%Type = Article
\bibitem[{Gomes~Jr et~al.(2023)Gomes~Jr, do~Lago, R{\'a}palo, Oliveira, Giacomoni and Mendiondo}]{gomes2023hydropol2d}
\bibinfo{author}{Gomes~Jr, M.N.}, \bibinfo{author}{do~Lago, C.A.F.}, \bibinfo{author}{R{\'a}palo, L.M.C.}, \bibinfo{author}{Oliveira, P.T.S.}, \bibinfo{author}{Giacomoni, M.H.}, \bibinfo{author}{Mendiondo, E.M.}, \bibinfo{year}{2023}.
\newblock \bibinfo{title}{Hydropol2d—distributed hydrodynamic and water quality model: Challenges and opportunities in poorly-gauged catchments}.
\newblock \bibinfo{journal}{Journal of Hydrology} , \bibinfo{pages}{129982}.
%Type = Article
\bibitem[{Gomes~Jr et~al.(2024c)Gomes~Jr, Taha, R{\'a}pallo, Mendiondo and Giacomoni}]{gomes2024real}
\bibinfo{author}{Gomes~Jr, M.N.}, \bibinfo{author}{Taha, A.F.}, \bibinfo{author}{R{\'a}pallo, L.M.C.}, \bibinfo{author}{Mendiondo, E.M.}, \bibinfo{author}{Giacomoni, M.H.}, \bibinfo{year}{2024}c.
\newblock \bibinfo{title}{Real-time regulation of detention ponds via feedback control: Balancing flood mitigation and water quality}.
\newblock \bibinfo{journal}{arXiv preprint arXiv:2403.04675} .
%Type = Article
\bibitem[{Green and Ampt(1911)}]{green1911studies}
\bibinfo{author}{Green, W.H.}, \bibinfo{author}{Ampt, G.}, \bibinfo{year}{1911}.
\newblock \bibinfo{title}{Studies on soil phyics.}
\newblock \bibinfo{journal}{The Journal of Agricultural Science} \bibinfo{volume}{4}, \bibinfo{pages}{1--24}.
%Type = Article
\bibitem[{Hu et~al.(2019)Hu, Lei, Han, Cao, Liu, He and Yue}]{hu2019improved}
\bibinfo{author}{Hu, P.}, \bibinfo{author}{Lei, Y.}, \bibinfo{author}{Han, J.}, \bibinfo{author}{Cao, Z.}, \bibinfo{author}{Liu, H.}, \bibinfo{author}{He, Z.}, \bibinfo{author}{Yue, Z.}, \bibinfo{year}{2019}.
\newblock \bibinfo{title}{Improved local time step for 2d shallow-water modeling based on unstructured grids}.
\newblock \bibinfo{journal}{Journal of Hydraulic Engineering} \bibinfo{volume}{145}, \bibinfo{pages}{06019017}.
%Type = Article
\bibitem[{Huff(1967)}]{huff1967time}
\bibinfo{author}{Huff, F.A.}, \bibinfo{year}{1967}.
\newblock \bibinfo{title}{Time distribution of rainfall in heavy storms}.
\newblock \bibinfo{journal}{Water resources research} \bibinfo{volume}{3}, \bibinfo{pages}{1007--1019}.
%Type = Article
\bibitem[{Hunter et~al.(2007)Hunter, Bates, Horritt and Wilson}]{hunter2007simple}
\bibinfo{author}{Hunter, N.M.}, \bibinfo{author}{Bates, P.D.}, \bibinfo{author}{Horritt, M.S.}, \bibinfo{author}{Wilson, M.D.}, \bibinfo{year}{2007}.
\newblock \bibinfo{title}{Simple spatially-distributed models for predicting flood inundation: A review}.
\newblock \bibinfo{journal}{Geomorphology} \bibinfo{volume}{90}, \bibinfo{pages}{208--225}.
%Type = Article
\bibitem[{Hunter et~al.(2005)Hunter, Horritt, Bates, Wilson and Werner}]{hunter2005adaptive}
\bibinfo{author}{Hunter, N.M.}, \bibinfo{author}{Horritt, M.S.}, \bibinfo{author}{Bates, P.D.}, \bibinfo{author}{Wilson, M.D.}, \bibinfo{author}{Werner, M.G.}, \bibinfo{year}{2005}.
\newblock \bibinfo{title}{An adaptive time step solution for raster-based storage cell modelling of floodplain inundation}.
\newblock \bibinfo{journal}{Advances in water resources} \bibinfo{volume}{28}, \bibinfo{pages}{975--991}.
%Type = Misc
\bibitem[{Husqvarna-Group(2021)}]{HUGSI}
\bibinfo{author}{Husqvarna-Group}, \bibinfo{year}{2021}.
\newblock \bibinfo{title}{Husqvarna urban green space index}.
\newblock \URLprefix \url{https://www.hugsi.green/city/?S\%C3\%A3o\%20Paulo}. \bibinfo{note}{accessed: 10-02-2021}.
%Type = Article
\bibitem[{Jones et~al.(2008)Jones, Sudicky and McLaren}]{jones2008application}
\bibinfo{author}{Jones, J.}, \bibinfo{author}{Sudicky, E.}, \bibinfo{author}{McLaren, R.}, \bibinfo{year}{2008}.
\newblock \bibinfo{title}{Application of a fully-integrated surface-subsurface flow model at the watershed-scale: A case study}.
\newblock \bibinfo{journal}{Water Resources Research} \bibinfo{volume}{44}.
%Type = Article
\bibitem[{Jongman et~al.(2012)Jongman, Ward and Aerts}]{jongman2012global}
\bibinfo{author}{Jongman, B.}, \bibinfo{author}{Ward, P.J.}, \bibinfo{author}{Aerts, J.C.}, \bibinfo{year}{2012}.
\newblock \bibinfo{title}{Global exposure to river and coastal flooding: Long term trends and changes}.
\newblock \bibinfo{journal}{Global Environmental Change} \bibinfo{volume}{22}, \bibinfo{pages}{823--835}.
%Type = Article
\bibitem[{Kim et~al.(2012)Kim, Warnock, Ivanov and Katopodes}]{kim2012coupled}
\bibinfo{author}{Kim, J.}, \bibinfo{author}{Warnock, A.}, \bibinfo{author}{Ivanov, V.Y.}, \bibinfo{author}{Katopodes, N.D.}, \bibinfo{year}{2012}.
\newblock \bibinfo{title}{Coupled modeling of hydrologic and hydrodynamic processes including overland and channel flow}.
\newblock \bibinfo{journal}{Advances in Water Resources} \bibinfo{volume}{37}, \bibinfo{pages}{104--126}.
%Type = Article
\bibitem[{do~Lago et~al.(2023)do~Lago, Giacomoni, Bentivoglio, Taormina, Junior and Mendiondo}]{do2023generalizing}
\bibinfo{author}{do~Lago, C.A.}, \bibinfo{author}{Giacomoni, M.H.}, \bibinfo{author}{Bentivoglio, R.}, \bibinfo{author}{Taormina, R.}, \bibinfo{author}{Junior, M.N.G.}, \bibinfo{author}{Mendiondo, E.M.}, \bibinfo{year}{2023}.
\newblock \bibinfo{title}{Generalizing rapid flood predictions to unseen urban catchments with conditional generative adversarial networks}.
\newblock \bibinfo{journal}{Journal of Hydrology} \bibinfo{volume}{618}, \bibinfo{pages}{129276}.
%Type = Article
\bibitem[{Liro et~al.(2022)Liro, Nones, Miku{\'s} and Plesi{\'n}ski}]{liro2022modelling}
\bibinfo{author}{Liro, M.}, \bibinfo{author}{Nones, M.}, \bibinfo{author}{Miku{\'s}, P.}, \bibinfo{author}{Plesi{\'n}ski, K.}, \bibinfo{year}{2022}.
\newblock \bibinfo{title}{Modelling the effects of dam reservoir backwater fluctuations on the hydrodynamics of a small mountain stream}.
\newblock \bibinfo{journal}{Water} \bibinfo{volume}{14}, \bibinfo{pages}{3166}.
%Type = Article
\bibitem[{Lu et~al.(2022)Lu, Xia and Shoemaker}]{lu2022surrogate}
\bibinfo{author}{Lu, W.}, \bibinfo{author}{Xia, W.}, \bibinfo{author}{Shoemaker, C.A.}, \bibinfo{year}{2022}.
\newblock \bibinfo{title}{Surrogate global optimization for identifying cost-effective green infrastructure for urban flood control with a computationally expensive inundation model}.
\newblock \bibinfo{journal}{Water Resources Research} \bibinfo{volume}{58}, \bibinfo{pages}{e2021WR030928}.
%Type = Article
\bibitem[{Luke et~al.(2015)Luke, Kaplan, Neal, Lant, Sanders, Bates and Alsdorf}]{luke2015hydraulic}
\bibinfo{author}{Luke, A.}, \bibinfo{author}{Kaplan, B.}, \bibinfo{author}{Neal, J.}, \bibinfo{author}{Lant, J.}, \bibinfo{author}{Sanders, B.}, \bibinfo{author}{Bates, P.}, \bibinfo{author}{Alsdorf, D.}, \bibinfo{year}{2015}.
\newblock \bibinfo{title}{Hydraulic modeling of the 2011 new madrid floodway activation: a case study on floodway activation controls}.
\newblock \bibinfo{journal}{Natural Hazards} \bibinfo{volume}{77}, \bibinfo{pages}{1863--1887}.
%Type = Article
\bibitem[{Marengo et~al.(2023)Marengo, Alcantara, Cunha, Seluchi, Nobre, Dolif, Goncalves, Dias, Cuartas, Bender et~al.}]{marengo2023flash}
\bibinfo{author}{Marengo, J.A.}, \bibinfo{author}{Alcantara, E.}, \bibinfo{author}{Cunha, A.P.}, \bibinfo{author}{Seluchi, M.}, \bibinfo{author}{Nobre, C.}, \bibinfo{author}{Dolif, G.}, \bibinfo{author}{Goncalves, D.}, \bibinfo{author}{Dias, M.A.}, \bibinfo{author}{Cuartas, L.}, \bibinfo{author}{Bender, F.}, et~al., \bibinfo{year}{2023}.
\newblock \bibinfo{title}{Flash floods and landslides in the city of recife, northeast brazil after heavy rain on may 25--28, 2022: Causes, impacts, and disaster preparedness}.
\newblock \bibinfo{journal}{Weather and Climate Extremes} \bibinfo{volume}{39}, \bibinfo{pages}{100545}.
%Type = Article
\bibitem[{Merwade et~al.(2008)Merwade, Olivera, Arabi and Edleman}]{merwade2008uncertainty}
\bibinfo{author}{Merwade, V.}, \bibinfo{author}{Olivera, F.}, \bibinfo{author}{Arabi, M.}, \bibinfo{author}{Edleman, S.}, \bibinfo{year}{2008}.
\newblock \bibinfo{title}{Uncertainty in flood inundation mapping: Current issues and future directions}.
\newblock \bibinfo{journal}{Journal of hydrologic engineering} \bibinfo{volume}{13}, \bibinfo{pages}{608--620}.
%Type = Article
\bibitem[{Mohanty et~al.(2020)Mohanty, Mudgil and Karmakar}]{mohanty2020flood}
\bibinfo{author}{Mohanty, M.P.}, \bibinfo{author}{Mudgil, S.}, \bibinfo{author}{Karmakar, S.}, \bibinfo{year}{2020}.
\newblock \bibinfo{title}{Flood management in india: A focussed review on the current status and future challenges}.
\newblock \bibinfo{journal}{International Journal of Disaster Risk Reduction} \bibinfo{volume}{49}, \bibinfo{pages}{101660}.
%Type = Article
\bibitem[{Muthusamy et~al.(2021)Muthusamy, Casado, Butler and Leinster}]{muthusamy2021understanding}
\bibinfo{author}{Muthusamy, M.}, \bibinfo{author}{Casado, M.R.}, \bibinfo{author}{Butler, D.}, \bibinfo{author}{Leinster, P.}, \bibinfo{year}{2021}.
\newblock \bibinfo{title}{Understanding the effects of digital elevation model resolution in urban fluvial flood modelling}.
\newblock \bibinfo{journal}{Journal of hydrology} \bibinfo{volume}{596}, \bibinfo{pages}{126088}.
%Type = Book
\bibitem[{Nachtergaele et~al.(2009)Nachtergaele, van Velthuizen, Verelst, Wiberg, Henry, Chiozza, Yigini, Aksoy, Batjes, Boateng et~al.}]{nachtergaele2023harmonized}
\bibinfo{author}{Nachtergaele, F.}, \bibinfo{author}{van Velthuizen, H.}, \bibinfo{author}{Verelst, L.}, \bibinfo{author}{Wiberg, D.}, \bibinfo{author}{Henry, M.}, \bibinfo{author}{Chiozza, F.}, \bibinfo{author}{Yigini, Y.}, \bibinfo{author}{Aksoy, E.}, \bibinfo{author}{Batjes, N.}, \bibinfo{author}{Boateng, E.}, et~al., \bibinfo{year}{2009}.
\newblock \bibinfo{title}{Harmonized World Soil Database version 2.0}.
\newblock \bibinfo{publisher}{Food and Agriculture Organization of the United Nations}.
%Type = Article
\bibitem[{Neal et~al.(2012a)Neal, Schumann and Bates}]{neal2012subgrid}
\bibinfo{author}{Neal, J.}, \bibinfo{author}{Schumann, G.}, \bibinfo{author}{Bates, P.}, \bibinfo{year}{2012}a.
\newblock \bibinfo{title}{A subgrid channel model for simulating river hydraulics and floodplain inundation over large and data sparse areas}.
\newblock \bibinfo{journal}{Water Resources Research} \bibinfo{volume}{48}.
%Type = Article
\bibitem[{Neal et~al.(2012b)Neal, Villanueva, Wright, Willis, Fewtrell and Bates}]{neal2012much}
\bibinfo{author}{Neal, J.}, \bibinfo{author}{Villanueva, I.}, \bibinfo{author}{Wright, N.}, \bibinfo{author}{Willis, T.}, \bibinfo{author}{Fewtrell, T.}, \bibinfo{author}{Bates, P.}, \bibinfo{year}{2012}b.
\newblock \bibinfo{title}{How much physical complexity is needed to model flood inundation?}
\newblock \bibinfo{journal}{Hydrological Processes} \bibinfo{volume}{26}, \bibinfo{pages}{2264--2282}.
%Type = Article
\bibitem[{Nithila~Devi and Kuiry(2024)}]{nithila2024novel}
\bibinfo{author}{Nithila~Devi, N.}, \bibinfo{author}{Kuiry, S.N.}, \bibinfo{year}{2024}.
\newblock \bibinfo{title}{A novel local-inertial formulation representing subgrid scale topographic effects for urban flood simulation}.
\newblock \bibinfo{journal}{Water Resources Research} \bibinfo{volume}{60}, \bibinfo{pages}{e2023WR035334}.
%Type = Article
\bibitem[{Patel et~al.(2017)Patel, Ramirez, Srivastava, Bray and Han}]{patel2017assessment}
\bibinfo{author}{Patel, D.P.}, \bibinfo{author}{Ramirez, J.A.}, \bibinfo{author}{Srivastava, P.K.}, \bibinfo{author}{Bray, M.}, \bibinfo{author}{Han, D.}, \bibinfo{year}{2017}.
\newblock \bibinfo{title}{Assessment of flood inundation mapping of surat city by coupled 1d/2d hydrodynamic modeling: a case application of the new hec-ras 5}.
\newblock \bibinfo{journal}{Natural Hazards} \bibinfo{volume}{89}, \bibinfo{pages}{93--130}.
%Type = Misc
\bibitem[{de~Pernambuco(2019)}]{PE3D}
\bibinfo{author}{de~Pernambuco, G.}, \bibinfo{year}{2019}.
\newblock \bibinfo{title}{Pe3d – pernambuco tridimensional. o que é o programa?}
\newblock \URLprefix \url{http://www.pe3d.pe.gov.br}.
%Type = Article
\bibitem[{Phyo et~al.(2023)Phyo, Yabar and Richards}]{phyo2023managing}
\bibinfo{author}{Phyo, A.P.}, \bibinfo{author}{Yabar, H.}, \bibinfo{author}{Richards, D.}, \bibinfo{year}{2023}.
\newblock \bibinfo{title}{Managing dam breach and flood inundation by hec-ras modeling and gis mapping for disaster risk management}.
\newblock \bibinfo{journal}{Case Studies in Chemical and Environmental Engineering} \bibinfo{volume}{8}, \bibinfo{pages}{100487}.
%Type = Misc
\bibitem[{PMSP(2017)}]{GeoSampa}
\bibinfo{author}{PMSP, S.P.P.m.d.S.P.}, \bibinfo{year}{2017}.
\newblock \bibinfo{title}{Portal geosampa}.
\newblock \URLprefix \url{http://geosampa.prefeitura.sp.gov.br/PaginasPublicas/_SBC.aspx#}. \bibinfo{note}{accessed: 12-04-2022}.
%Type = Article
\bibitem[{Pontes et~al.(2017)Pontes, Fan, Fleischmann, de~Paiva, Buarque, Siqueira, Jardim, Sorribas and Collischonn}]{pontes2017mgb}
\bibinfo{author}{Pontes, P.R.M.}, \bibinfo{author}{Fan, F.M.}, \bibinfo{author}{Fleischmann, A.S.}, \bibinfo{author}{de~Paiva, R.C.D.}, \bibinfo{author}{Buarque, D.C.}, \bibinfo{author}{Siqueira, V.A.}, \bibinfo{author}{Jardim, P.F.}, \bibinfo{author}{Sorribas, M.V.}, \bibinfo{author}{Collischonn, W.}, \bibinfo{year}{2017}.
\newblock \bibinfo{title}{Mgb-iph model for hydrological and hydraulic simulation of large floodplain river systems coupled with open source gis}.
\newblock \bibinfo{journal}{Environmental Modelling \& Software} \bibinfo{volume}{94}, \bibinfo{pages}{1--20}.
%Type = Article
\bibitem[{R{\'a}palo et~al.(2024)R{\'a}palo, Gomes~Jr and Mendiondo}]{rapalo4703477developing}
\bibinfo{author}{R{\'a}palo, L.}, \bibinfo{author}{Gomes~Jr, M.N.}, \bibinfo{author}{Mendiondo, E.M.}, \bibinfo{year}{2024}.
\newblock \bibinfo{title}{Developing an open-source flood forecasting system adapted to data-scarce regions: A digital twin coupled with hydrologic-hydrodynamic simulations}.
\newblock \bibinfo{journal}{Available at SSRN 4703477} .
%Type = Article
\bibitem[{Reshma et~al.(2024)Reshma, Devi and Kuiry}]{reshma2024real}
\bibinfo{author}{Reshma, R.}, \bibinfo{author}{Devi, N.N.}, \bibinfo{author}{Kuiry, S.N.}, \bibinfo{year}{2024}.
\newblock \bibinfo{title}{Real-time urban flood modeling: exploring the sub-grid approach for accurate simulation and hazard analysis}.
\newblock \bibinfo{journal}{Natural Hazards} , \bibinfo{pages}{1--39}.
%Type = Article
\bibitem[{Sanders(2008)}]{sanders2008integration}
\bibinfo{author}{Sanders, B.F.}, \bibinfo{year}{2008}.
\newblock \bibinfo{title}{Integration of a shallow water model with a local time step}.
\newblock \bibinfo{journal}{Journal of Hydraulic Research} \bibinfo{volume}{46}, \bibinfo{pages}{466--475}.
%Type = Article
\bibitem[{Schumann et~al.(2013)Schumann, Neal, Voisin, Andreadis, Pappenberger, Phanthuwongpakdee, Hall and Bates}]{schumann2013first}
\bibinfo{author}{Schumann, G.P.}, \bibinfo{author}{Neal, J.C.}, \bibinfo{author}{Voisin, N.}, \bibinfo{author}{Andreadis, K.M.}, \bibinfo{author}{Pappenberger, F.}, \bibinfo{author}{Phanthuwongpakdee, N.}, \bibinfo{author}{Hall, A.C.}, \bibinfo{author}{Bates, P.D.}, \bibinfo{year}{2013}.
\newblock \bibinfo{title}{A first large-scale flood inundation forecasting model}.
\newblock \bibinfo{journal}{Water Resources Research} \bibinfo{volume}{49}, \bibinfo{pages}{6248--6257}.
%Type = Article
\bibitem[{Sharifian et~al.(2022)Sharifian, Kesserwani, Chowdhury, Neal and Bates}]{sharifian2022lisflood}
\bibinfo{author}{Sharifian, M.K.}, \bibinfo{author}{Kesserwani, G.}, \bibinfo{author}{Chowdhury, A.A.}, \bibinfo{author}{Neal, J.}, \bibinfo{author}{Bates, P.}, \bibinfo{year}{2022}.
\newblock \bibinfo{title}{Lisflood-fp 8.1: New gpu accelerated solvers for faster fluvial/pluvial flood simulations}.
\newblock \bibinfo{journal}{Geoscientific Model Development Discussions} \bibinfo{volume}{2022}, \bibinfo{pages}{1--28}.
%Type = Article
\bibitem[{Siqueira et~al.(2018)Siqueira, Paiva, Fleischmann, Fan, Ruhoff, Pontes, Paris, Calmant and Collischonn}]{siqueira2018toward}
\bibinfo{author}{Siqueira, V.A.}, \bibinfo{author}{Paiva, R.C.}, \bibinfo{author}{Fleischmann, A.S.}, \bibinfo{author}{Fan, F.M.}, \bibinfo{author}{Ruhoff, A.L.}, \bibinfo{author}{Pontes, P.R.}, \bibinfo{author}{Paris, A.}, \bibinfo{author}{Calmant, S.}, \bibinfo{author}{Collischonn, W.}, \bibinfo{year}{2018}.
\newblock \bibinfo{title}{Toward continental hydrologic--hydrodynamic modeling in south america}.
\newblock \bibinfo{journal}{Hydrology and Earth System Sciences} \bibinfo{volume}{22}, \bibinfo{pages}{4815--4842}.
%Type = Misc
\bibitem[{SIURB(2020)}]{SIURB}
\bibinfo{author}{SIURB}, \bibinfo{year}{2020}.
\newblock \bibinfo{title}{Manchas de inundação}.
\newblock \URLprefix \url{https://geosampa.prefeitura.sp.gov.br/PaginasPublicas/downloadArquivo.aspx?orig=DownloadCamadas&arq=10_Meio\%20F\%EDsico\%5C\%5C\%C1rea\%20Inund\%E1vel\%5C\%5CShapefile\%5C\%5CSIRGAS_SHP_areainundavel&arqTipo=Shapefile}. \bibinfo{note}{accessed: 12-04-2024}.
%Type = Article
\bibitem[{Sridharan et~al.(2020)Sridharan, Gurivindapalli, Kuiry, Mali, Nithila~Devi, Bates and Sen}]{sridharan2020explicit}
\bibinfo{author}{Sridharan, B.}, \bibinfo{author}{Gurivindapalli, D.}, \bibinfo{author}{Kuiry, S.N.}, \bibinfo{author}{Mali, V.K.}, \bibinfo{author}{Nithila~Devi, N.}, \bibinfo{author}{Bates, P.D.}, \bibinfo{author}{Sen, D.}, \bibinfo{year}{2020}.
\newblock \bibinfo{title}{Explicit expression of weighting factor for improved estimation of numerical flux in local inertial models}.
\newblock \bibinfo{journal}{Water Resources Research} \bibinfo{volume}{56}, \bibinfo{pages}{e2020WR027357}.
%Type = Article
\bibitem[{Towsif~Khan et~al.(2020)Towsif~Khan, Chapa and Hack}]{towsif2020highly}
\bibinfo{author}{Towsif~Khan, S.}, \bibinfo{author}{Chapa, F.}, \bibinfo{author}{Hack, J.}, \bibinfo{year}{2020}.
\newblock \bibinfo{title}{Highly resolved rainfall-runoff simulation of retrofitted green stormwater infrastructure at the micro-watershed scale}.
\newblock \bibinfo{journal}{Land} \bibinfo{volume}{9}, \bibinfo{pages}{339}.
%Type = Article
\bibitem[{Tschiedel and Paiva(2018)}]{tschiedel2018uncertainty}
\bibinfo{author}{Tschiedel, A.d.F.}, \bibinfo{author}{Paiva, R.C.D.d.}, \bibinfo{year}{2018}.
\newblock \bibinfo{title}{Uncertainty assessment in hydrodynamic modeling of floods generated by dam break}.
\newblock \bibinfo{journal}{RBRH} \bibinfo{volume}{23}, \bibinfo{pages}{e30}.
%Type = Article
\bibitem[{Tschiedel et~al.(2020)Tschiedel, Paiva and Fan}]{tschiedel2020use}
\bibinfo{author}{Tschiedel, A.d.F.}, \bibinfo{author}{Paiva, R.C.D.d.}, \bibinfo{author}{Fan, F.M.}, \bibinfo{year}{2020}.
\newblock \bibinfo{title}{Use of large-scale hydrological models to predict dam break-related impacts}.
\newblock \bibinfo{journal}{RBRH} \bibinfo{volume}{25}, \bibinfo{pages}{e35}.
%Type = Article
\bibitem[{Wang et~al.(2021)Wang, Chen and Huang}]{wang2021urban}
\bibinfo{author}{Wang, W.}, \bibinfo{author}{Chen, W.}, \bibinfo{author}{Huang, G.}, \bibinfo{year}{2021}.
\newblock \bibinfo{title}{Urban stormwater modeling with local inertial approximation form of shallow water equations: A comparative study}.
\newblock \bibinfo{journal}{International Journal of Disaster Risk Science} \bibinfo{volume}{12}, \bibinfo{pages}{745--763}.
%Type = Book
\bibitem[{Weschenfelder et~al.(2020)Weschenfelder, Pickbrenner and Pinto}]{Atlas2020}
\bibinfo{author}{Weschenfelder, A.B.}, \bibinfo{author}{Pickbrenner, K.}, \bibinfo{author}{Pinto, E.J.d.A.}, \bibinfo{year}{2020}.
\newblock \bibinfo{title}{Atlas Pluviométrico do Brasil: Equações Intensidade-Duração-Frequência (Desagregação de Precipitações Diárias); município: Tietê; Estação Pluviométrica: Tietê, Código 02347056}.
\newblock \bibinfo{address}{Porto Alegre}.
%Type = Article
\bibitem[{Xing et~al.(2022)Xing, Chen, Liang and Ma}]{xing2022improving}
\bibinfo{author}{Xing, Y.}, \bibinfo{author}{Chen, H.}, \bibinfo{author}{Liang, Q.}, \bibinfo{author}{Ma, X.}, \bibinfo{year}{2022}.
\newblock \bibinfo{title}{Improving the performance of city-scale hydrodynamic flood modelling through a gis-based dem correction method}.
\newblock \bibinfo{journal}{Natural Hazards} , \bibinfo{pages}{1--23}.
%Type = Article
\bibitem[{Yamazaki et~al.(2013)Yamazaki, de~Almeida and Bates}]{yamazaki2013improving}
\bibinfo{author}{Yamazaki, D.}, \bibinfo{author}{de~Almeida, G.A.}, \bibinfo{author}{Bates, P.D.}, \bibinfo{year}{2013}.
\newblock \bibinfo{title}{Improving computational efficiency in global river models by implementing the local inertial flow equation and a vector-based river network map}.
\newblock \bibinfo{journal}{Water Resources Research} \bibinfo{volume}{49}, \bibinfo{pages}{7221--7235}.
%Type = Article
\bibitem[{Yazdi(2019)}]{yazdi2019optimal}
\bibinfo{author}{Yazdi, J.}, \bibinfo{year}{2019}.
\newblock \bibinfo{title}{Optimal operation of urban storm detention ponds for flood management}.
\newblock \bibinfo{journal}{Water Resources Management} \bibinfo{volume}{33}, \bibinfo{pages}{2109--2121}.

\end{thebibliography}




% \end{thebibliography}
\end{document}